\documentclass[pra, aps, twocolumn,amssymb,amsmath,amsbsy, groupedaddress, superscriptaddress]{revtex4-1}
\usepackage{slashed}
\usepackage{graphicx}
\usepackage{float}
\usepackage{subcaption}
\usepackage{enumerate}
\usepackage{chngcntr}
\usepackage[usenames, dvipsnames]{color}
\usepackage{graphics,amsfonts,amsmath, amssymb}
\usepackage{mathtools}
\usepackage{hyperref}
\usepackage{bm}
\usepackage{amsfonts}
\usepackage{dsfont}
\usepackage{lipsum}
\usepackage[utf8]{inputenc}
\usepackage[T1]{fontenc}
\usepackage{bbold}
\usepackage{times}
\usepackage{color}

\usepackage[squaren,Gray]{SIunits}
\hypersetup{backref,
colorlinks=true,
urlcolor=blue,
linkcolor=blue,
linktoc=page,
citecolor=blue}

\newcommand{\bra}[1]{\langle{#1}|}
\newcommand{\ket}[1]{|{#1}\rangle}

\everymath{\displaystyle} 

\begin{document}
	
	\title{\Large Enhanced multiparameter quantum estimation in cavity magnomechanics via a coherent feedback loop}
	\date{\today}
	\author{Adnan Naimy}
	\affiliation{LPHE, Modeling et Simulations, Faculty of Science, Mohammed V University in Rabat, Morocco.}
	\author{Abdallah Slaoui}
	\affiliation{LPHE, Modeling et Simulations, Faculty of Science, Mohammed V University in Rabat, Morocco.}
	\affiliation{CPM, Centre of Physics and Mathematics, Faculty of Science, Mohammed V University in Rabat, Morocco.}\affiliation{Center of Excellence in Quantum and Intelligent Computing, Prince Sultan University, Riyadh, Saudi Arabia.}\author{Abderrahim Lakhfif}
	\affiliation{LPHE, Modeling et Simulations, Faculty of Science, Mohammed V University in Rabat, Morocco.}
	\affiliation{CPM, Centre of Physics and Mathematics, Faculty of Science, Mohammed V University in Rabat, Morocco.}\author{Rachid Ahl Laamara}
	\affiliation{LPHE, Modeling et Simulations, Faculty of Science, Mohammed V University in Rabat, Morocco.}
	\affiliation{CPM, Centre of Physics and Mathematics, Faculty of Science, Mohammed V University in Rabat, Morocco.}
\begin{abstract}
Multiparameter quantum metrology plays a fundamental role in uncovering and exploiting the distinctive features of quantum systems. In this work, we propose an effective and experimentally feasible scheme to significantly enhance the simultaneous quantum estimation of the photon–magnon and magnon–mechanical coupling strengths in a hybrid cavity–magnon–mechanical platform. Our approach relies on the assistance of a coherent feedback loop combined with the injection of a coherent driving field. We show that an appropriate tuning of the system and feedback parameters leads to a substantial reduction of the estimation errors associated with both coupling strengths. To quantify the metrological performance of the proposed scheme, we employ the quantum Cramér–Rao bound (QCRB) as a fundamental benchmark for multiparameter estimation. We explicitly compute and compare the QCRBs derived from the symmetric logarithmic derivative (SLD) and the right logarithmic derivative (RLD) formalisms. Our results demonstrate that the RLD-based QCRB is systematically lower than the SLD-based bound, indicating superior estimation precision in the considered noncommutative estimation scenario. We further analyze the performance of heterodyne detection and show that, in suitable parameter regimes, the corresponding classical estimation precision closely approaches the ultimate quantum limit predicted by our scheme. Finally, we discuss the experimental feasibility of the proposed setup within currently available cavity–magnon–mechanical platforms. Owing to its general character, the framework developed here can be readily extended to the high-precision estimation of other physical parameters in hybrid quantum systems.

\end{abstract}

\maketitle
\section{INTRODUCTION}
The primary objective of quantum parameter estimation theory is to determine the fundamental precision limits for measuring unknown parameters that characterize quantum systems, as well as to identify practical strategies capable of reaching these limits. Quantum metrology thus aims to develop methods that exploit quantum resources in order to determine the ultimate precision bounds imposed by quantum estimation theory. This approach enables the design of new quantum sensors that meet the growing demand for highly efficient and cost-effective detection technologies. Single-parameter quantum estimation has been extensively studied and applied to develop quantum strategies that enhance the measurement of various physical quantities. When a large number of independent repetitions is available, the attainable precision limit is given by the quantum Cramér–Rao bound (QCRB), originally introduced by Helstrom \cite{Helstrom1969,Holevo2011,Braunstein1994}. This bound represents the minimum achievable variance in the estimation of a parameter and can typically be saturated within the framework of single-parameter estimation.\par 

However, when multiple parameters are estimated simultaneously, saturating the QCRB becomes more challenging due to the incompatibility of the optimal measurements associated with each parameter \cite{Matsumoto2002,Vidrighin2014,Crowley2014,Ragy2016}. This difficulty has motivated the development of multiparameter quantum metrology, which aims to generalize the conditions under which the QCRB can be saturated and thus achieve maximal precision. In many practical scenarios, it is necessary to estimate several parameters at once \cite{Szczykulska2016,Albarelli2020,Demkowicz2020}, such as the relative positions of optical sources \cite{Napoli2019,Bisketzi2019}, the moments of extended sources \cite{Zhou2019,Tsang2019}, phase and noise parameters \cite{Crowley2014,Pinel2013,Altorio2015,Roccia2018,Szczykulska2017,Jayakumar2024}, the distance and velocity of moving targets \cite{Reichert2024,Huang2021}, or orthogonal displacements \cite{Yuen2003,Bradshaw2018,Park2022,Hanamura2023,Zhou2025,Frigerio2025}.\par
In general, determining the QCRB requires the calculation of the quantum Fisher information matrix (QFIM) \cite{Safranek2018-1,Gao2014}. This matrix plays a central role in multiparameter quantum metrology, as its inverse defines the fundamental precision limits for simultaneous parameter estimation. Enhancing the QFIM is therefore a key challenge for improving multiparameter estimation protocols. The QFIM appears in several research fields, including quantum thermometry \cite{Correa2015,Hofer2017}, gravitational-wave detection \cite{Abbott2016}, the estimation of squeezing parameters \cite{Milburn1994,Chiribella2006}, and the study of relativistic effects such as the Unruh–Hawking phenomenon \cite{Aspachs2010}.\\
To evaluate the QFIM, particularly in continuous-variable (CV) systems described by Gaussian states, various analytical techniques have been developed. Gaussian states have attracted considerable interest in quantum information science \cite{Ferraro2005,Braunstein2005} owing to two major advantages: their experimental accessibility and their analytical tractability. These states find applications in numerous domains, including quantum teleportation protocols \cite{Wolf2007} and cavity optomechanics \cite{Nunnenkamp2011,Peng2025}. In the latter context, the authors of \cite{Naimy2025} applied multiparameter estimation theory to an optomechanical cavity system, where they estimated two physical parameters with significant impact on the system’s behavior.\par
Motivated by this line of research, we consider a cavity magnomechanical system with coherent feedback to estimate two key coupling strengths of the system: the magnon–cavity coupling $g_{ma}$ and the magnomechanical coupling $g_{md}$. The cavity magnomechanical platform consists of a Fabry–Perot cavity integrated with a YIG (yttrium–iron garnet) sphere. YIG-based magnons have attracted increased attention due to their remarkable properties \cite{Huebl2012,Tabuchi2014,Xiong2023}, including low damping rates, high spin density, and broad tunability. The Kittel mode \cite{Kittel1948} in a YIG sphere can realize strong coupling with microwave photons in a high-quality cavity, giving rise to cavity polaritons \cite{Tabuchi2014,Bai2015} and vacuum Rabi splitting. In cavity magnomechanics, a magnon mode (spin wave) couples simultaneously to a vibrational deformation mode of the ferromagnet via magnetostrictive forces, and to a microwave cavity mode through magnetic dipole interactions. The magnetostrictive interaction is dispersive and analogous to radiation-pressure coupling in optomechanics, particularly when the mechanical frequency is much smaller than the magnon frequency \cite{Zhang2016,Fan2022}.\\
On the other hand, coherent feedback—where the system output is reinjected coherently without a measurement stage—offers a powerful means of controlling quantum systems while avoiding measurement-induced noise \cite{Lloyd2000,Nelson2000}. Recent theoretical and experimental works have confirmed its effectiveness for mechanical cooling \cite{Huang2019,Ernzer2023}, enhanced estimation of the optomechanical coupling \cite{Peng2025}, and the generation of squeezed and stationary entangled states in optomechanical and magnomechanical platforms \cite{Harwood2020,Bemani2024,Li2017,Peng2023,Zheng2023}.\par
Following the experimental realization of our cavity magnomechanical setup, we extract the fundamental statistical quantities—the covariance matrix and the displacement vector—from the measured output data. These two quantities fully characterize the Gaussian state of the system and serve as essential inputs for advanced quantum-metrology protocols. Indeed, they contain all the necessary quantum information for computing optimal estimation metrics such as the QFIM and the corresponding QCRB. Using these fundamental inputs, we proceed with the metrological protocol to evaluate the ultimate sensitivities achievable for estimating the coupling parameters of the system.\par
The main motivation behind introducing such a hybrid scheme is to determine which method—based on the symmetric logarithmic derivative (SLD) or on the right logarithmic derivative (RLD)—provides superior precision for estimating the two coupling strengths. The role of coherent feedback is to investigate the enhancement of the QCRB and identify conditions under which it becomes significantly improved.\par
The scheme considered in this work corresponds to an experimentally implemented setup designed for the simultaneous estimation of two distinct parameters in a cavity-based magnomechanical system incorporating coherent feedback. A microwave field is used to enhance the magnon–photon coupling. At the level of the YIG sphere, three mutually orthogonal magnetic fields are applied: a bias field along $z$, the cavity-mode magnetic field along $x$, and the pump field along $y$.\\
In our system, the couplings $g_{ma}$ and $g_{md}$ play a central role in the distribution and transfer of quantum information between the modes. The parameter $g_{ma}$ governs the interaction between the magnon mode and the cavity field, determining the efficiency of correlation transfer and the generation of optomagnonic correlations. The coupling $g_{md}$, in contrast, regulates the energy and information exchange between the magnonic and mechanical modes, impacting mechanical coherence and dynamical behavior. Joint optimization of these couplings maximizes the available Fisher information for parameter estimation while mitigating the effects of losses and thermal noise.\par
Understanding the impact of $g_{ma}$ and $g_{md}$ is therefore essential for developing effective coherent-feedback strategies and achieving high-precision measurements in hybrid magnomechanical platforms. To evaluate the ultimate precision limits, we analyze the QFIM by deriving its general analytical expression using both RLD \cite{Holevo2003} and SLD \cite{Safranek2018} approaches. The phase-space formalism proves particularly powerful in this context, as it greatly simplifies the calculation of key quantities in multiparameter quantum estimation and yields analytical results suitable for continuous-variable systems.

This paper is organized as follows. In Sec. \ref{Gaussian Multiparameter Estimation Theory}, we introduce the theoretical framework for multiparameter estimation of Gaussian states. We derive the expressions of the QFIMs associated with the SLD and the RLD using the vectorization method. We also evaluate the classical Fisher information (CFI) obtained from heterodyne detection. In Sec. \ref{Model Description and Dynamics in Cavity Magnonics}, we present the hybrid cavity–magnonics system under consideration, introduce its Hamiltonian, and derive the corresponding Heisenberg–Langevin equations that govern the system dynamics. These equations are then employed to determine the steady-state covariance matrix. In Sec. \ref{APPLICATION}, we apply the formalism developed in the previous sections to analyze the influence of key system parameters on the estimation of the cavity–magnon and magnon–mechanical coupling strengths, with particular emphasis on the role of the coherent feedback loop. In Sec. \ref{Experimental feasibility} we discuss the experimental feasibility of the proposed scheme. Finally, Sec. \ref{CONCLUSIONS} summarizes the main findings and concludes the paper.

\begin{widetext}

\begin{figure}[H]
    \centering
    \includegraphics[width=0.9\textwidth]{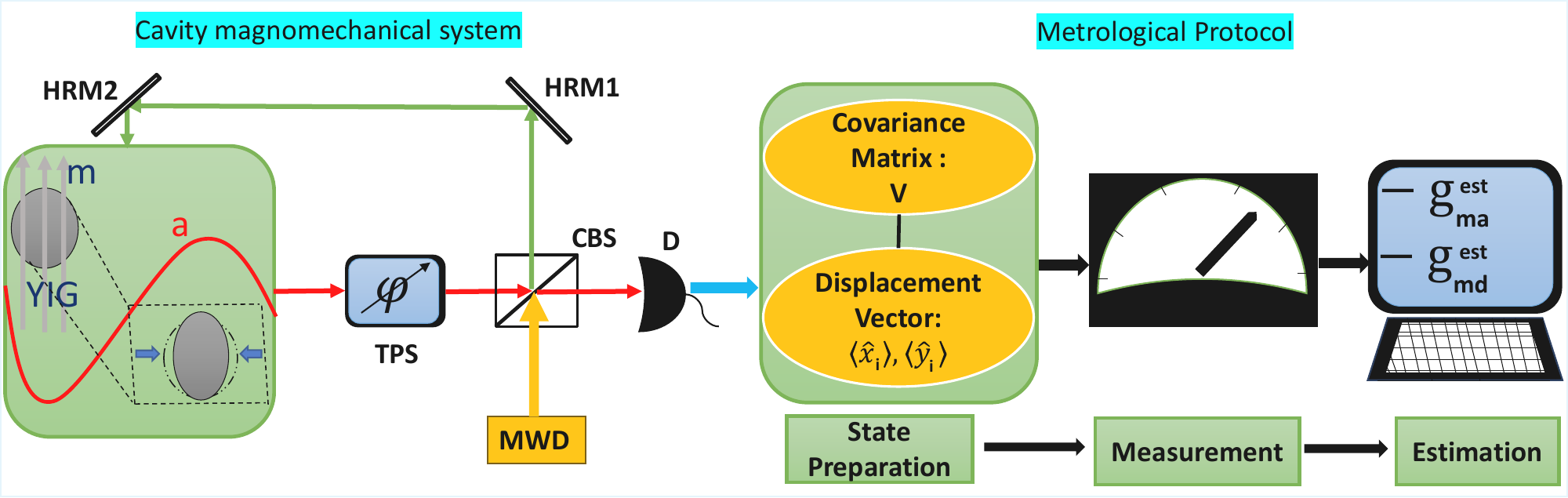}
    \caption{Schematic of a hybrid cavity–magnonics system with coherent feedback. A microwave driving field (MWD) is injected into a controlled beam splitter (CBS), characterized by reflection and transmission coefficients $r$ and $\varepsilon$, respectively. The transmitted component propagates through two highly reflective mirrors (HRM1 and HRM2) before entering a single-port microwave cavity. Inside the cavity, a single-mode electromagnetic field coherently couples to a magnon mode, realized by a YIG sphere positioned near the maximum of the cavity magnetic field and subjected to a uniform bias magnetic field. The cavity output field is directed to a detection setup, from which the first-order moments (quadratures) and the covariance matrix are extracted, enabling a full reconstruction of the Gaussian state of the system. This reconstructed state serves as the prepared resource for metrological applications. Based on the measurement outcomes of the output field, a statistical estimator is constructed, allowing the inference of the parameters of interest and the assessment of the estimation precision.} \label{shema met and mag}
\end{figure}
\end{widetext}
\section{Gaussian Multiparameter Estimation Theory} \label{Gaussian Multiparameter Estimation Theory}
In this section, we first present the main mathematical tools and definitions required to determine the fundamental limits of multiparameter quantum estimation protocols. We then introduce the formalism used to describe continuous-variable Gaussian quantum systems. In addition, we employ the heterodyne detection method to derive the classical Fisher information, which can be applied to the estimation of any physical parameter in our system.\\
Quantum parameter estimation theory investigates the ultimate precision limits achievable in determining physical parameters encoded in quantum states, as well as the types of measurements that can saturate these limits. In this context, we consider the simultaneous estimation of $n$ unknown parameters $\theta=\{\theta_1,\theta_2,...,\theta_n\}$, which characterize a parametric family of quantum states $\Tilde{\rho}_\theta$. These states describe the evolution of a quantum system governed by a Hamiltonian that depends on the parameters to be estimated, forming what is known as a quantum statistical model. The parameters $\theta$, assumed to belong to an open subset of $\mathrm{R^n}$, are incorporated into the system’s dynamics through an evolution map $\varrho_\theta$. To estimate them, one performs a general quantum measurement: a probe state $\rho$ is first prepared and then evolves according to $\rho \xrightarrow{\varrho_{\boldsymbol{\theta}}} \rho_{\boldsymbol{\theta}}$. The measurement is mathematically described by a positive operator-valued measure (POVM) $\{\text{M}_x\}$. The probability of obtaining a measurement outcome $x$ is given by Born’s rule; $\mathcal{P}(x/\varphi) = Tr[\hat{\rho}_{\theta}\;\hat{\text{M}}_x]$, where $\hat{\text{M}}_x$ denotes the POVM element associated with the outcome $x$. Based on the observed result, an estimator function $\hat{\theta}(x)$ is used to infer the parameter values. The performance of this estimator is quantified by the mean square error matrix (MSEM), defined as, $\Sigma_\theta\hat{\theta}(x)=\int dx \mathcal{P}(x/\theta) \left(\hat{\theta}(x)-\theta\right)\left(\hat{\theta}(x)-\theta\right)^T$. The quantum Cramér–Rao bound (QCRB) then sets the fundamental and physically attainable limit for any quantum estimation strategy. It defines the best achievable precision in evaluating unknown parameters of a quantum state and thus serves as a cornerstone of quantum metrology. In general, this bound can be written as follows
\begin{align}
   \text{ Cov}(\hat{\theta}) \ge & \frac{1}{W}\, (\mathrm{F}^S)^{-1}, \label{covS}\\
    \text{ Cov}(\hat{\theta}) \ge & \frac{1}{W}\, \text{Re}[(\mathrm{F}^R)^{-1}]+\big|\big|\text{Im}[(\mathrm{F}^R)^{-1}]\big|\big|_1, \label{covR}
\end{align}

where $W$ denotes the number of measurement repetitions, $\big|\big|  A\big|\big|_1$ is the trace norm of the matrix $A$, and $ \text{ Cov}(\hat{\theta})$ represents the covariance matrix, defined as $ \text{ Cov}(\theta_k, \theta_l)=E(\theta_k\, \theta_l)-E(\theta_k)E(\theta_l)$. The matrices $\mathrm{F^S}$ and $\mathrm{F^R}$ correspond to the QFIMs associated with the SLD and the RLD, respectively, and are expressed as follows
\begin{align}
     \mathrm{F}^S_{k,l} =& \;\text{Re} \left(Tr\big[\hat{L}^S_k\hat{\rho}_\theta \hat{L}^S_l\big]\right), \nonumber\\
     =&\, \frac{1}{2} Tr\big[\hat{\rho}_\theta\{\hat{L}^S_{l}, \hat{L}^S_{k}\}\big], \hspace{0.5cm}(SLD) \label{F SLD}\\
      \mathrm{F}_{k,l}^R =&\, Tr\big[ \hat{L}^{R,\dagger}_k\, \hat{\rho}_\theta\; \hat{L}^{{R}}_{l}\big].\hspace{0.5cm}(RLD)\label{F RLD}
\end{align}
where $L^S$ and $L^R$ denote the symmetric and right logarithmic derivative (SLD and RLD) operators, respectively, which are defined as the solutions of the following equations \cite{Braunstein1994,Genoni2013,Gao2014,Chang2025}
\begin{align}
    \{\hat{\rho}_\theta,\,\hat{L}_k^S\}=&\,2\frac{\partial\hat{\rho}_\theta}{\partial\theta_k}, \hspace{0.5cm}(SLD) \label{SLD} \\
    \hat{\rho}_\theta\,\hat{L}_k^R=&\frac{\partial \hat{\rho}_\theta}{\partial\theta_k}.\hspace{0.5cm}(RLD)\label{RLD}
\end{align}

The curly brackets $\{.,.\}$ denote the anticommutator. The quantity $\text{Re}(Tr[\hat{A}\,\hat{\rho}\,\hat{B}])$ is known as the SLD inner product between two Hermitian operators $\hat{A}$ and $\hat{B}$. The SLD operators are Hermitian and have zero mean, i.e.,
$Tr[\hat{\rho}_\theta\,\hat{L}^S_k]=0$, since the partial derivatives $\partial\hat{\rho}_\theta/\partial\theta_k$ have vanishing trace due to the normalization condition of $\rho_\theta$. On the other hand, the RLD operators are generally non-Hermitian but also have zero mean, $Tr[\hat{\rho}_\theta\,\hat{L}^R_k]=0$. The quantity $Tr[\hat{A}^\dagger\,\hat{\rho}\,\hat{B}]$ represents the RLD inner product between the operators $\hat{A}$ and $\hat{B}$, which in this case are not necessarily Hermitian. The quantum Fisher information matrix associated with the SLD (SLD-QFIM) is real, symmetric, and positive definite. To evaluate the performance of a quantum estimation protocol, it is customary to quantify the estimation precision through the variance of the estimated parameters. This serves as a meaningful figure of merit for assessing the efficiency of multiparameter estimation strategies and for comparing different theoretical bounds. Consequently, Eqs. \eqref{covS} and \eqref{covR} can be rewritten in a scalar form that directly characterizes the attainable precision limits,
\begin{align}
    \text{Var}(\theta_k) \ge& \frac{1}{W}\, (\mathrm{F}^S_{\theta_k \theta_k})^{-1}, \label{varFS}\\
     \text{Var}(\theta_k) \ge& \frac{1}{W}\,\left( \text{Re}\left((\mathrm{F^R}_{\theta_k \theta_k})^{-1}\right)+\big|\big|\text{Im}\left((\mathrm{F^R}_{\theta_k \theta_k})^{-1}\right)\big|\big|_1\right). \label{varFR}
\end{align}
The two preceding equations are always saturated, indicating that the parameter estimation reaches the ultimate precision allowed by quantum mechanics. Such saturation occurs when the measurement is performed in the eigenbasis of the SLD operators, with the optimal states corresponding to the projectors onto their eigenvectors. By taking the trace over these relations, one obtains explicit expressions for the complete set of estimated parameters, as presented below
\begin{align}
     \sum_k^n \text{Var}(\theta_k) \ge& \,\mathrm{C}^S=\frac{1}{W}\, Tr\left[(\mathrm{F}^S)^{-1}\right], \label{TrFS}\\
     \sum_k^n\text{Var}(\theta_k) \ge& \,\mathrm{C}^R=\frac{1}{W}\,\left( \text{Re}\left((\mathrm{F^R})^{-1}\right)+\big|\big|\text{Im}\left((\mathrm{F^R})^{-1}\right)\big|\big|_1\right). \label{TrFR}
\end{align}

In most practical situations, neither the RLD bound nor the SLD bound can be exactly saturated \cite{Monras2011}. This limitation originates from the intrinsically non-commutative nature of quantum mechanics, which prevents achieving optimal precision when estimating multiple parameters simultaneously. In practice, improving the precision for one parameter typically degrades the accuracy attainable for the others. It is also important to note that the estimator saturating the RLD bound is not always associated with a physically implementable POVM. However, in scenarios where the optimal measurements for each individual parameter are mutually non-commuting, it remains possible to saturate both the RLD and SLD bounds by employing a suitably engineered joint-measurement strategy. A substantial portion of the research in multiparameter quantum estimation has predominantly focused on the SLD-based bound, as discussed in \cite{Monras2011,Monras2010,Crowley2014}, with further developments presented in \cite{Genoni2013}. In the single-parameter regime, it is well established that the SLD-QFI is always less than or equal to the RLD-QFI, thereby providing a stricter bound on sensitivity \cite{Helstrom1969,Monras2011}. In the multiparameter setting, recent advances based on the theory of quantum local asymptotic normality (QLAN) have demonstrated that the matrix bound in Eqs. \eqref{covS} and \eqref{varFS} becomes asymptotically achievable under appropriate conditions,
\begin{equation}
Tr\left[\hat{\rho}\left[\hat{\mathrm{L}}^S_{\theta_k}, \hat{\mathrm{L}}^S_{\theta_l}\right]\right] = 0. \label{cond rho Ls}
\end{equation}
The latter condition, following the discussion in Ref. \cite{Nichols2018}, can be equivalently written as
\begin{equation}
    \mathcal{I}m \left(Tr\left[\hat{\rho}\, \hat{\mathrm{L}}^{S}_{\theta_k}\, \hat{\mathrm{L}}^{S}_{\theta_l}\right]\right)=0.
\end{equation}

Failure to satisfy the above condition implies that the bound 
$\mathrm{C}^R$ (associated with the RLD) becomes more relevant than the bound $\mathrm{C}^S$ (associated with the SLD). Nevertheless, obtaining explicit analytical forms for the RLD and SLD operators, as well as for their associated QFIM, generally remains a challenging task—except in specific cases where the density matrix can be readily diagonalized \cite{Aspachs2010,Monras2007,Zhang2013}. It is also important to emphasize that the two bounds, $\mathrm{C}^S$
(associated with the SLD) and $\mathrm{C}^R$ (associated with the RLD), differ significantly, particularly in their mathematical structure, as each arises from a distinct derivation procedure. In this work, we apply the established expressions for the SLD and RLD to a magnomechanical system under coherent feedback in order to compute the corresponding QCRBs. This naturally raises the question of which bound provides the most informative characterization of the estimation precision.\\
To address this issue, the authors of Refs. \cite{Genoni2013,Gao2014,Naimy2025} introduced a quantity defined as the minimum between the two bounds derived from the SLD and RLD. This quantity, referred to as the Most Informative Quantum Cramér–Rao Bound $\mathrm{C}^{MI}$, is given by
\begin{equation}
    \mathrm{C}^{MI} = \text{Min} \{\mathrm{C}^S,\mathrm{C}^R \},\label{CMI}
\end{equation}
where
\begin{align}
    \mathrm{C}^S=&\,\frac{1}{W}.\frac{\mathrm{F^S_{\theta_k \theta_k}}+\mathrm{F^S_{\theta_l \theta_l}}}{\text{det}(\mathrm{F}^S)},\\
    \mathrm{C}^R=&\,\frac{1}{W}.\frac{\mathcal{R}e(\mathrm{F^R_{\theta_k \theta_k}})+\mathcal{R}e(\mathrm{F^R_{\theta_l \theta_l}})}{\;\mathcal{R}e(\text{det}(\mathrm{F}^R))}\nonumber\\
    &+\frac{1}{W}.\frac{||\mathcal{I}m(\mathrm{F^R_{\theta_k \theta_k}})||_1+||\mathcal{I}m(\mathrm{F^R_{\theta_l \theta_l}})||_1}{\mathcal{I}m(\text{det}(\mathrm{F}^R))},
\end{align}
with one obtains the determinant of the QFIM as \\$\text{det}(\mathrm{F})=\,\mathrm{F}_{\theta_k \theta_k}.\mathrm{F}_{\theta_l \theta_l}-\mathrm{F}_{\theta_k \theta_l}.\mathrm{F}_{\theta_l \theta_k}$. In order to determine which of the two bounds, $\mathrm{C}^R$ or $\mathrm{C}^S$, yields the more stringent precision constraint, we examine their ratio $\mathrm{C}^R/\mathrm{C}^S$. This ratio typically manifests in one of three distinct regimes, each characterizing a different comparative relationship between the RLD and SLD bounds. These scenarios provide crucial insight into which bound delivers the most informative precision limit based on specific system characteristics. More precisely, when $\mathrm{C}^R/\mathrm{C}^S>1$, the SLD bound $\mathrm{C}^S$ provides the more informative constraint; conversely, when $\mathrm{C}^R/\mathrm{C}^S<1$, the RLD bound $\mathrm{C}^R$ constitutes the more limiting precision bound. In the singular case where $\mathrm{C}^R/\mathrm{C}^S=1$, both bounds converge to identical precision limits, establishing the equivalence $\mathrm{C}^{MI}=\mathrm{C}^R=\mathrm{C}^S$. Based on Eq. \eqref{CMI}, one can assert that the optimal measurements in multiparameter estimation protocols are entirely determined by the $\mathrm{C}^{MI}$. Consequently, this bound satisfies the following inequality
\begin{equation}
    \sum_k^n \text{Var}(\theta_k) \ge \, \frac{1}{W}\, \mathrm{C}^{MI}.
\end{equation}

\subsection{QUANTUM ESTIMATION THEORY}
Quantum estimation theory provides a fundamental framework for determining the ultimate precision limits in the estimation of physical parameters in quantum systems. At the center of this framework lies the QCRB, which can be formulated using either the SLD or the RLD operators. To derive explicit analytical expressions for these operators and their associated QFIMs, we rely on the vectorization technique, introduced in the subsection below.\par
\textbf{Vectorization Method:} Diagonalizing the density matrix is a common approach to derive an analytical expression for the QFIM. In our system, however, we work with the displacement vector and the covariance matrix, and for this purpose, we employ an alternative technique known as vectorization \cite{Safranek2018,Chang2025}. This method has proven to be highly effective and is applicable to any finite-dimensional covariance matrix. It relies on the use of the Vec operator, which allows one to transform a matrix into a column vector by stacking its columns. Specifically, for any matrix $\mathrm{A}\in \mathcal{M}^{n\times n}$, where $\mathcal{M}^{n\times n}$ denotes the space of $n\times n$ matrices, the operator $\text{Vec}[\mathrm{A}]$ is defined as follows
\begin{equation}
        \mathrm{A}  = \begin{pmatrix}
        a_{11}  & \cdots & a_{1n}  \\
        \vdots  & \ddots & \vdots  \\
        a_{n1}  & \cdots & a_{nn}  
    \end{pmatrix} \longmapsto \text{Vec}[\mathrm{A}]= \begin{pmatrix}
        a_{11}    \\
        \vdots    \\
        a_{n1}   \\
        \vdots    \\
        a_{1n}    \\
        \vdots    \\
        a_{nn}
    \end{pmatrix}
\end{equation}
The following property linking the vectorization operator with Kronecker products will be employed in the subsequent derivations
\begin{equation}
    \text{Vec}[ABC] =\left(C^{T}\otimes A\right)  \text{Vec}[B]
\end{equation}
The vectorization procedure, commonly used for Hilbert-space operators in finite-dimensional systems, finds a distinct application here. Our approach consists of applying it to matrices characterizing the phase-space description of continuous-variable systems, specifically the covariance matrix and the matrix defining the quadratic component of operators. \par
\textbf{Assessment of the RLD-QFIM:} To derive the RLD operators, we employ a somewhat technical approach that has already been developed in Ref. \cite{Gao2014}, albeit using a different notation. We present the corresponding results here for completeness and to highlight the formal analogy with the SLD case. For a Gaussian state, the RLD operator $\hat{\mathrm{L}}^R_{\theta_k}$ in Eq. \eqref{RLD} is at most quadratic in the canonical operators and can therefore be written in the following form
\begin{equation}
    \hat{\mathrm{L}}^R_{\theta k}=\; \hat{\mathrm{L}}^{R(0)} + \hat{\mathrm{L}}^{R(1)}_1 \hat{{\text{R}}}_1 + \hat{\mathrm{L}}^{R(2)}_{23} \hat{{\text{R}}}_2 \hat{{\text{R}}}_3.\label{LR}
\end{equation}
where $\hat{{\text{R}}}$ denotes the vector of canonical operators, $\hat{\mathrm{L}}^{R(0)} \in \mathbb{C}$, $\hat{\mathrm{L}}^{R(1)} \in \mathbb{C}^{2N}$, $\hat{\mathrm{L}}^{R(2)} \in \mathbb{C}^{2N\times2N}$, with $\mathbb{C}^{2N}$ and $\mathbb{C}^{2N\times2N}$, referring respectively to the sets of complex vectors and complex matrices.
For a given set of parameters $\theta_k$, the quantities $\hat{\mathrm{L}}^{R(0)}_{\theta_k}$, $\hat{\mathrm{L}}^{R(1)}_{\theta_k}$ and $\hat{\mathrm{L}}^{R(2)}_{\theta_k}$ appearing in Eq. \eqref{LR} can be expressed, respectively, as follows
\begin{subequations}
\begin{align}
  \hat{\mathrm{L}}^{R(0)}_{\theta_k} =& -\frac{1}{2}Tr\left[(2\mathrm{V}+i\Omega)_+  \hat{\mathrm{L}}^{R(2)}_{\theta_k} \right] - \langle R \rangle^\mathbf{T}  \hat{\mathrm{L}}^{R(1)}_{\theta_k}\nonumber \\
  &- \langle R \rangle^\mathbf{T}  \hat{\mathrm{L}}^{R(2)}_{\theta_k}\langle R \rangle,\\
  \hat{\mathrm{L}}^{R(1)}_{\theta_k} =& \,2 (2\mathrm{V}+i\Omega)_+^{-1} \partial_{\theta_k} \langle R \rangle - 2 \hat{\mathrm{L}}^{R(2)}_{\theta_k}\langle R \rangle,\\
  \text{Vec} \left[ \hat{\mathrm{L}}^{R(2)}_{\theta_k} \right] =& \left((2\mathrm{V}+i\Omega)^\dagger \otimes (2\mathrm{V}+i\Omega) \right)^+ \text{Vec} \left[\partial_{\theta_k} \mathrm{V} \right], 
\end{align} 
\end{subequations}
where $\mathrm{V}$ denotes the covariance matrix, $\langle R \rangle$ is the displacement vector, and $\Omega=\oplus_{k=1}^n i\,\sigma_y$, with $\sigma_y$ being the second Pauli matrix. By substituting the obtained expressions of the RLD operator into the general formula of the RLD-QFIM given in Eq. \eqref{F RLD}, we arrive at
\begin{align}
     \mathrm{F}_{\theta_k\theta_l}^R =& \;2\, \text{vec}\left[\partial_{\theta_k}\mathrm{V} \right]^\dagger \left((2\mathrm{V}+i\Omega)^\dagger\otimes(2\mathrm{V}+i\Omega)\right)^+  \text{Vec}\left[\partial_{\theta_l}\mathrm{V} \right] \nonumber\\
     &+ 2 \partial_{\theta_k}  \langle R \rangle^{\mathbf{T}} (2\mathrm{V}+i\Omega)^{+} \partial_{\theta_l}  \langle R \rangle,\label{FR with psedo}
\end{align}
where the symbol '$+$' refers to the Moore–Penrose pseudoinverse, which extends the usual notion of matrix inversion \cite{Penrose1955,Ben-Israel2003}. This pseudoinverse can be evaluated through Tikhonov regularization \cite{Golub1996}, namely: $\text{A}^+=\lim\limits_{x \rightarrow 0} \big(\text{A}^\dagger(\text{A}\text{A}^\dagger+ \delta I)^{-1}\big)=\lim\limits_{x \rightarrow 0} \big((\text{A}\text{A}^\dagger+ \delta I)^{-1}\text{A}^\dagger\big)$. These limits remain well-defined even when the ordinary inverse $A^{-1}$
 does not exist. Moreover, when the matrices $(2\mathrm{V}+i\Omega)$ and $(2\mathrm{V}+i\Omega)\otimes(2\mathrm{V}+i\Omega)$ are nonsingular —so that their Moore–Penrose pseudoinverses reduce to the standard matrix inverses— the RLD-QFIM admits the following simplified form
 \begin{align}
     \mathrm{F}_{\theta_k\theta_l}^R =& \;2\, \text{vec}\left[\partial_{\theta_k}\mathrm{V} \right]^\dagger \left((2\mathrm{V}+i\Omega)^\dagger\otimes(2\mathrm{V}+i\Omega)\right)^{-1}  \text{Vec}\left[\partial_{\theta_l}\mathrm{V} \right] \nonumber\\
     &+ 2 \partial_{\theta_k}  \langle R \rangle^{\mathbf{T}} (2\mathrm{V}+i\Omega)^{-1} \partial_{\theta_l}  \langle R \rangle,\label{FR with inv}
\end{align}
\textbf{Assessment of the SLD-QFIM:} Gaussian states possess a remarkable property: their SLD operators are at most quadratic polynomials in the canonical operators. This characteristic has been extensively leveraged in the literature to derive precise analytical expressions for these operators and for the ensuing QFIM \cite{Nichols2018,Gao2014,Monras2013,Serafini2023,Safranek2018}. For the sake of completeness, we recall these key results here. Starting from the working ansatz that the SLD is a quadratic function of the canonical operators, it can be expressed in the standard basis through the general form given in Eq. \eqref{LR}
\begin{equation}
    \hat{\mathrm{L}}^S_{\theta k}=\; \hat{\mathrm{L}}^{S(0)} + \hat{\mathrm{L}}^{S(1)}_1 \hat{{\text{R}}}_1 + \hat{\mathrm{L}}^{S(2)}_{23} \hat{{\text{R}}}_2 \hat{{\text{R}}}_3.\label{LS}
\end{equation}
where the quantities $\hat{\mathrm{L}}^{S(0)} \in \mathbb{R}$, $\hat{\mathrm{L}}^{S(1)} \in \mathbb{R}^{2N}$, and $\hat{\mathrm{L}}^{S(2)} \in \mathbb{R}^{2N\times2N}$ are a scalar, a vector, and a matrix, respectively. Here, $\mathbb{R}^{2N}$ and $\mathbb{R}^{2N\times2N}$ denote the sets of real vectors and real matrices of dimensions $2N$ and $2N\times2N$, respectively. These quantities are given by
\begin{subequations}
\begin{align}
 &\hat{\mathrm{L}}^{S(0)}_{\theta_k} =-Tr\left[\mathrm{V}\;  \hat{\mathrm{L}}^{S(2)}_{\theta_k} \right] - \langle R \rangle^\mathbf{T}  \hat{\mathrm{L}}^{S(1)}_{\theta_k} - \langle R \rangle^\mathbf{T}  \hat{\mathrm{L}}^{S(2)}_{\theta_k}\langle R \rangle , \\
&\hat{\mathrm{L}}^{S(1)}_{\theta_k} = \mathrm{V}^{-1} \partial_{\theta_k} \langle R \rangle - 2 \hat{\mathrm{L}}^{S(2)}_{\theta_k}\langle R \rangle,\\
&\text{vec} \left[ \hat{\mathrm{L}}^{S(2)}_{\theta_k} \right] = \left(4\mathrm{V}^\dagger \otimes \mathrm{V} + \Omega \otimes \Omega \right)^+ \text{vec} \left[\partial_{\theta_k} \mathrm{V} \right] .
\end{align} 
\end{subequations}
Therefore, the elements of the SLD-QFIM are obtained by substituting the SLD expression into Eq. \eqref{F SLD}.
\begin{align}
     \mathrm{F}^S_{\theta_k \theta_l} =& \, 2\,\text{Vec}\left[\partial_{\theta_k}\mathrm{V} \right]^\dagger  
     \left(4\mathrm{V}^\dagger \otimes \mathrm{V} + \Omega \otimes \Omega \right)^{+}  \text{Vec}\left[\partial_{\theta_l}\mathrm{V} \right] \nonumber\\
     &+  \partial_{\theta_k}  \langle R \rangle^{\mathbf{T}} \mathrm{V}^{-1} \partial_{\theta_l}  \langle R \rangle.\label{FS with inv}
\end{align}
If the matrix $\left(4\mathrm{V}^\dagger \otimes \mathrm{V} + \Omega \otimes \Omega \right)$ is invertible, the SLD-QFIM is given by the following new expression
\begin{align}
     \mathrm{F}^S_{\theta_k \theta_l} =& \, 2\,\text{Vec}\left[\partial_{\theta_k}\mathrm{V} \right]^\dagger  
     \left(4\mathrm{V}^\dagger \otimes \mathrm{V} + \Omega \otimes \Omega \right)^{-1}  \text{Vec}\left[\partial_{\theta_l}\mathrm{V} \right] \nonumber\\
     &+  \partial_{\theta_k}  \langle R \rangle^{\mathbf{T}} \mathrm{V}^{-1} \partial_{\theta_l}  \langle R \rangle.\label{FS with inv}
\end{align}

\subsection{CLASSICAL ESTIMATION THEORY}
In this section, we adopt a measurement strategy based on heterodyne detection, which is particularly well suited for parameter estimation in our hybrid system. This measurement technique is widely recognized for its effectiveness in Gaussian quantum systems, as it allows simultaneous access to both field quadratures and enables an optimization of the estimation precision. The primary goal of this approach is to reduce the mean squared error (MSE) associated with the estimation of unknown parameters, while allowing— in the case of unbiased estimators — the attainment or near saturation of the Cramér–Rao bound (CRB).

From a formal perspective, parameter estimation is performed through a general quantum measurement described by a set of positive-operator-valued measures (POVMs), denoted by $\{\hat{\text{M}}_x\}$, \big[with $x=(x_1,x_2,...)^T, \, \text{M}_x \ge0,\; \int dx\,\text{M}_x=\mathbb{1}$\big] \cite{Gao2014}. Each measurement operator $\hat{\text{M}}_x$ satisfies the completeness relation $\sum_x \hat{\text{M}}_x^\dagger \hat{\text{M}}_x =\mathbb{1}$ \cite{Degen2017,Giovannetti2011,Giovannetti2006,Liu2020}, thereby ensuring a consistent and physically admissible description of the quantum measurement process. Once the measurement scheme is fixed, the quantum system yields a conditional probability distribution $\mathcal{P}(x/\theta)$, which explicitly depends on the unknown parameter $\theta$ to be estimated. This probability distribution plays a central role in quantifying the estimation precision, as it allows the introduction of the classical Fisher information (CFI). The CFI characterizes the sensitivity of the measurement statistics to variations of the parameter of interest and determines the ultimate precision achievable with the chosen measurement strategy. It is explicitly defined as follows \cite{Chang2025}
\begin{equation}
    \mathrm{F}^C_{\theta_k \theta_l} = \int \mathcal{P}(x/\theta) \left[\frac{\partial \,\text{ln}\, \mathcal{P}(x/\theta)}{\partial\theta_k}\right]\left[\frac{\partial \,\text{ln}\,\mathcal{P}(x/\theta)}{\partial\theta_l}\right]dx.\label{CFI Ge}
\end{equation}

The conditional probability describes the distribution of measurement outcomes obtained in the estimation of the parameter $x$ after the measurement process. Its expression is given by the Born rule and depends explicitly on the unknown parameter $\theta$, as follows
\begin{equation}
    \mathcal{P}(x/\theta) = Tr[\rho_{\theta}\;\text{M}_x].
\end{equation}
Rather than relying solely on the CFI to estimate the system parameters and assess the performance of the measurement strategy, it is also appropriate to introduce the classical Cramér–Rao bound (CRB). This bound establishes a fundamental lower limit on the mean squared error (MSE) achievable by any unbiased estimator. In particular, when an estimator saturates the CRB, it is regarded as efficient, as it fully exploits the information contained in the measurement data. In general, the CRB is expressed by the following inequality
\begin{equation}
    \text{Cov}(\hat{\theta})=\sum_k^n \text{Var}(\theta_k)\ge\mathrm{C}^C=\frac{1}{W\,\mathrm{F}^C_{\theta_k}}
\end{equation}
where $\mathrm{C}^C$ denotes the CRB. In general, this bound is constrained by the QCRB $\mathrm{C}^Q$, which establishes the ultimate precision limit allowed by quantum mechanics. The equality $\mathrm{C}^C=\mathrm{C}^Q$ can only be achieved under specific conditions, requiring the implementation of an optimal measurement strategy. Such an optimal strategy is commonly associated with the SLD, from which the corresponding classical Fisher information matrix (CFIM) can be constructed. The SLD operator $\hat{L}_\theta$ is implicitly defined through the relation $2\,\partial\hat{\rho}_\theta=\hat{\rho}\hat{L}_\theta+\hat{\rho}\hat{L}_\theta$. However, despite its theoretical optimality, this measurement configuration may be difficult to realize experimentally and, in some cases, may even be inaccessible. This indicates that the maximum estimation precision predicted by the QCRB is not always attainable in practical implementations.

In this work, we focus on the CRB obtained via heterodyne detection. We further demonstrate that this classical bound can closely approach the enhanced QCRB $\mathrm{C}^C\ge\mathrm{C}^Q$ generated by our system. Heterodyne detection constitutes a key measurement strategy for Gaussian states, in which the field under investigation is mixed with a reference field of a different frequency. This technique enables simultaneous access to both optical quadratures which is particularly advantageous for multiparameter estimation.

In contrast to homodyne detection, which measures only a single quadrature at a time, heterodyne detection provides more complete information about the optical state. Moreover, it exhibits reduced sensitivity to phase fluctuations of the local oscillator and is widely employed in experimental platforms \cite{Pontin2018,Buonanno2003}. Formally, heterodyne detection is described by projection operators onto coherent states, given by $[\ket{\alpha}\bra{\alpha}/\pi]$ \cite{Aspelmeyer2014}. Based on Eq. \eqref{CFI Ge}, this formulation allows one to explicitly evaluate the CFI associated with heterodyne measurements, as detailed in Refs. \cite{Monras2013,Chang2025}.

\begin{equation}
\mathrm{F}_{\theta_k \theta_l}^C = \frac{1}{2} \operatorname{Tr}\left[ \mathcal{V}^{-1} \frac{\partial \mathcal{V}}{\partial \theta_k} \mathcal{V}^{-1} \frac{\partial \mathcal{V}}{\partial \theta_l} \right] + \left( \frac{\partial \langle R \rangle}{\partial \theta_k} \right)^T \mathcal{V}^{-1} \left(\frac{\partial \langle R \rangle}{\partial \theta_l}\right).\label{CF}
\end{equation}

The covariance matrix appearing in the above expression can be written in terms of the original covariance matrix that characterizes the system [see Eq. \eqref{matrix V}]. Specifically, it is given by $\mathcal{V}=\mathrm{V}+\mathbb{1}_{6\times6}$, where $\mathbb{1}_{6\times6}$ denotes the $6\times6$ identity matrix.

\section{Model Description and Dynamics in Cavity Magnonics}\label{Model Description and Dynamics in Cavity Magnonics}

We consider a hybrid cavity–magnonics platform as illustrated in Fig. \ref{shema met and mag}. The system consists of a one-sided microwave cavity embedding a $ 250-\mu m$-diameter YIG sphere and incorporating a coherent feedback loop in order to enhance its performance. The cavity is driven by an external microwave field with amplitude E and frequency $\omega_L$. The feedback loop comprises two highly reflective mirrors (HRM1 and HRM2) and a controlled beam splitter (CBS), characterized by transmission and reflection coefficients $\varepsilon$ and $r$, respectively, satisfying $\varepsilon^2+r^2=1$. A portion of the cavity output field is routed back to the cavity input through the loop, establishing coherent feedback, while the remaining part is transmitted toward the detection stage. The microwave cavity supports a single electromagnetic mode of frequency $\omega_a$, represented by the bosonic annihilation (creation) operators $a (a^\dagger)$. This mode interacts coherently with a magnon mode of frequency $\omega_m$, described by the operators $m (m^\dagger)$. Magnons correspond to quantized spin-wave excitations of the electron spin ensemble in the YIG sphere, positioned near the maximum of the cavity magnetic-field profile and simultaneously subjected to a uniform static magnetic field. The latter determines the magnon frequency through $\omega_m=\gamma B_0$, and ensures a strong magnon–photon coupling with strength $g_{ma}$ \cite{Zhang2014,Potts2020}. We take the size of the YIG sphere to be much smaller than the microwave wavelength, such that radiation-pressure effects can be safely neglected. In a frame rotating at the driving frequency $\omega_L$, the Hamiltonian of the system is given by
\begin{align}
    \mathcal{H}/\hbar=& \, \omega_{a} a^\dagger a + \omega_{m} m^\dagger m + \frac{\omega_d}{2}(q^2+p^2) + J_{md}\;m^+mq \nonumber\\
    &+g_{ma} (m+m^\dagger)(a+a^\dagger) +i  \Omega\big(m^+e^{-i\omega_L \textit{t}}-me^{i\omega_L \textit{t}} \big)\nonumber \\
    &+E \:\varepsilon (a e^{-i\varphi}+a^+ e^{i\varphi}). \label{Hamiltonien}
\end{align}

The first two terms of the system Hamiltonian describe the free evolution of the optical and magnonic modes. Here, $a (a^\dagger)$ and $m (m^\dagger)$ denote the annihilation (creation) operators of the cavity photon and magnon modes, respectively, while $\omega_a$ and $\omega_m$ are their corresponding resonance frequencies. The magnon frequency $\omega_m = \gamma H$ is controlled by the gyromagnetic ratio $\gamma$ and the external static magnetic field $H$. The next contribution represents the mechanical mode, characterized by the dimensionless position and momentum quadratures $q$ and $p$, which satisfy the canonical commutation relation $[q,p]=i$. Its resonance frequency is denoted by $\omega_d$. The magnon–mechanical interaction is described through a single-magnon magnomechanical coupling rate $J_{md}$, which is intrinsically weak; however, it can be significantly enhanced by strongly driving the magnon mode with a microwave field applied directly to the YIG sphere \cite{Wang2018}. The following term accounts for the coherent coupling between the microwave magnon mode and the cavity electromagnetic mode, with the coupling strength $g_{ma}$. In the strong-coupling regime typically realized in cavity magnonics, $g_{ma}$ exceeds the dissipation rates of both the cavity field $\kappa_a$ and the magnon mode $\kappa_m$, i.e. 
$g_{ma}> \kappa_a, \kappa_m$ \cite{Huebl2013,Tabuchi2014,Goryachev2014,Bai2015}. The magnon mode is additionally driven by an external coherent microwave source, characterized by a driving frequency $\omega_L$ and a Rabi frequency $\Omega$, which quantifies the interaction strength between the drive magnetic field and the magnon mode. The latter is related to the drive amplitude $B_0$ and to the total number of spins $N=\rho V$, where $\rho$ is the spin density of the YIG and $V$ the volume of the sphere. The expression for $\Omega$ is obtained under the standard low-excitation condition $⟨m^\dagger m⟩<< 2Ns$, with $s=5/2$ corresponding to the spin quantum number of $Fe^{3+}$ ions in YIG placed at the antinode of the magnetic field inside the cavity. The final term in the Hamiltonian represents the coherent optical drive transmitted to the cavity through the beam splitter, where $E=\sqrt{2\kappa_aP/\hbar \omega_L}$ denotes the drive amplitude and $\phi$ its phase. The rotating-wave approximation (RWA) applied to the cavity–magnon interaction, reducing $g_{ma} (m+m^\dagger)(a+a^\dagger)\to g_{ma} (m a^\dagger + m^\dagger a)$, is fully justified in this configuration. Indeed, both $\omega_a$ and $\omega_m$ lie in the microwave regime and are much larger than $g_{ma}$ and the corresponding decay rates. As a consequence, counter-rotating terms oscillate rapidly and average out over time \cite{Li2018,Zhang2016}. To describe the quantum dynamics of the hybrid cavity–magnon–mechanical system, we rely on the Heisenberg–Langevin formalism, which consistently incorporates damping mechanisms and quantum fluctuations. Starting from the Hamiltonian in Eq. \eqref{Hamiltonien}, we derive the quantum Langevin equations (QLEs) for all modes in the absence of coherent feedback i.e., by setting $\varepsilon=0$, which corresponds to removing the last term in the Hamiltonian. In a frame rotating at the driving frequency, the HLEs read
\begin{subequations}
    \begin{align}
    \frac{d\,q}{dt}=& \;\omega_d\,p,\\
        \frac{d\,p}{dt}=& - J_{md} \,m^\dagger m -\omega_d\,q-\gamma_d\,p  + \zeta ,\\
        \frac{d\,a}{dt} =& -(i\Delta_a+\kappa_a)a-ig_{ma}\,m+\sqrt{2\kappa_a} \,\eta_a,\\
        \frac{d\,m}{dt} =& -(i\Delta_m+\kappa_m)m -ig_{ma}\,c - i\,J_{md}\,mq  \nonumber\\ 
        &+ \Omega+\sqrt{2\kappa_m} \,\eta_m,
    \end{align}\label{HLEs}
\end{subequations}
here, $\Delta_a=\omega_a-\omega_L$ denote the detunings of the cavity and magnon modes from the driving field. The parameters 
$\kappa_a$ and $\kappa_m$ correspond to the decay rates of the cavity and magnon modes, while $\gamma_d$ represents the mechanical damping rate. The interaction of the mechanical mode with its thermal environment is modeled through the noise operator $\zeta$, which satisfies the non-vanishing correlation functions; $\langle \zeta(\textit{t})\,\zeta(\textit{s})+ \zeta(\textit{s})\,\zeta(\textit{t})\rangle/2 =\,\gamma_d\left[2 \,\text{n}(\omega_d)+1\right] \,\delta(\textit{t}-\textit{s})$. Similarly, the input noise operators of the cavity and magnon modes, denoted by $\eta_a$ and $\eta_m$, obey the standard Markovian correlations; $\langle \eta_j(\textit{t})\;\eta^{\dagger}(\textit{s})\rangle =\left[ \textit{n}(\omega_j)+1\right] \,\delta(\textit{t}-\textit{s}),$ $\langle \eta^{\dagger}(\textit{t})\; \eta(\textit{s})\rangle = \textit{n}(\omega_j) \,\delta(\textit{t}-\textit{s}),\;\{j\in  a,m\}$, where $\textit{n}(\omega_j)$ is the mean thermal occupancy of photons, magnons, or phonons, defined by  $\textit{n}(\omega_j)=\left(\text{exp}[\hbar \omega_j/k_BT]-1\right)^{-1},\; \{j\in  a,m,d\}$, with $\hbar$ the reduced Planck constant, $k_B$ the Boltzmann constant, and $T$ the environmental temperature \cite{Gardiner2004,Li2018}.

When coherent feedback is introduced, the effective input field driving the cavity, denoted $\eta_{a,fb}$, arises from the superposition of the direct microwave driving field $\eta_a$ and the portion of the cavity output field that is redirected back through the feedback loop. As a result, the input signal is modified into a coherent combination of the original drive and the reflected cavity output. Under these conditions, the HLE corresponding to the cavity mode is therefore rewritten in the following form
\begin{equation}
    \frac{d\,a}{dt} = -(i\Delta_{a,fb}+\kappa_{a,fb})a -ig_{ma}\,m-i\varepsilon E \;e^{i\varphi}+\sqrt{2\kappa_a} \,\Tilde{\eta}_a.\label{HLE,fb}
\end{equation}
The effective cavity detuning and decay rate in the presence of coherent feedback are given by $\Delta_{a,fb}=\Delta_a-2\,r\,\kappa_a \sin(\theta)$ and $\kappa_{a,fb} = \kappa_a-2\,r\,\kappa_a \cos(\theta)$, respectively, where $\theta$ denotes the phase shift introduced by the reflection of the cavity output field along the feedback loop. The magnon and mechanical HLEs in Eqs. \eqref{HLEs} are unchanged. In general, when coherent feedback is applied, the modified input field driving the cavity takes the form
\begin{equation}
    \eta_{a,fb}=r\,e^{i\theta}\eta_{out}(t-\tau)+\varepsilon\,\eta_a(t),
\end{equation}
where $\tau$ represents the round-trip delay time in the loop. The feedback can be treated as instantaneous by setting $\tau \to 0$, and by using the standard input–output relation $\eta_{out}(t)=-\varepsilon\,\eta_a(t)+\sqrt{2\kappa_a}\,a(t)$ The instantaneous-feedback approximation holds when $\tau \ll 1/\kappa_a$, ensuring that the signal propagation time in the external loop is negligible compared to the intrinsic response time of the system. In cavity magnonic platforms, $\kappa_a$ typically lies in the MHz range, confirming the validity of this condition. Under these assumptions, the effective feedback-modified input field at the cavity becomes
\begin{align}
    \eta_{a,fb} =& \;r\,e^{i\theta}\eta_{out}(t)+\varepsilon\,\eta_a(t)\nonumber \\
    =&\;\Tilde{\eta}_a(t) + r\,e^{i\theta}\sqrt{2\kappa_a}\,a(t),
\end{align}
where $\Tilde{\eta_a}(t)= \varepsilon(1-r\,e^{i\theta})\,\eta_a(t)$ denotes the cavity noise operator modified by the feedback. Its non-vanishing correlation functions are given by
\begin{subequations}    
\begin{align}
    \langle \eta_j(\textit{t})\;\eta^{\dagger}(\textit{s})\rangle =&\;\varepsilon^2 |1-r\,e^{i\theta}|^2 \left[ \textit{n}(\omega_a)+1\right] \,\delta(\textit{t}-\textit{s}),\\
    \langle \eta^{\dagger}(\textit{t})\; \eta(\textit{s})\rangle 
    =& \;\varepsilon^2 |1-r\,e^{i\theta}|^2\textit{n}(\omega_a) \,\delta(\textit{t}-\textit{s}).
\end{align}
\end{subequations}

To investigate the quantum dynamics of the cavity–magnon hybrid system, we linearize the HLEs [Eqs. \eqref{HLEs} and \eqref{HLE,fb}] around their steady-state values. This approximation is well justified when the coherent microwave drive $E$ is sufficiently strong, allowing the annihilation operators $a$ and $m$ to be expressed as the sum of their steady-state coherent amplitudes ($\alpha_s$ and $\beta_s$) and small zero-mean quantum fluctuations ($\delta a$) and ($\delta m$). By introducing the decompositions $a=\alpha_s+\delta a$ and $m=\beta_s+\delta m$ into the dynamical equations and neglecting second-order fluctuation terms, we derive the linearized HLEs governing the evolution of the quantum fluctuations around the steady state, as follows

\begin{subequations}
    \begin{align}
    \frac{d\,\delta q}{dt}=& \;\omega_d\,\delta p,\\
        \frac{d\,\delta p}{dt}=&  -\omega_d\,\delta q-\gamma_d\,\delta p -g_{md} \delta x  + \zeta ,\\
        \frac{d\,\delta a}{dt} =& -(i\Delta_{a,fb}+\kappa_{a,fb})\delta a-ig_{ma}\,\delta m+\sqrt{2\kappa_a} \,\Tilde{\eta}_a,\\
        \frac{d\,\delta m}{dt} =& -(i\Tilde{\Delta}_m+\kappa_m)\delta m -ig_{ma}\,\delta a - g_{md}\,\delta q  \nonumber\\ 
        &+\sqrt{2\kappa_m} \,\eta_m.
    \end{align}\label{HLEs-lineire}
\end{subequations}
To determine the steady-state expectation values $q_s$, $p_s$, $\alpha_s$, and $\beta_s$, we set the time derivatives in the HLEs to zero while ignoring the zero-mean noise contributions. This leads to
\begin{subequations}
    \begin{align}
       q_s=&\frac{-J_{md}|\beta_s|^2}{\omega_d},\\
       p_s=&0,\\
       \alpha_s=&-\frac{i\,g_{ma}\beta_s+i\varepsilon\, E e^{i\varphi}}{i\Delta_{a,fb}+\kappa_{a,fb}},\\
       \beta_s=&\frac{-ig_{ma}\alpha_s+\Omega}{i\Tilde{\Delta}_m+\kappa_m},
   \end{align}
\end{subequations}   
where $\Tilde{\Delta}_m=\Delta_m+J_{md}\,q_s$ represents the effective detuning of the magnon mode, incorporating the frequency shift induced by the magno-mechanical interaction, and $g_{md}=i\sqrt{2} J_{md}\,\beta_s $ denotes the effective magno-mechanical coupling strength.
The linearized HLEs in Eqs. \eqref{HLEs-lineire} can be expressed in a compact quadrature form as
\begin{equation}
    \frac{dN}{dt} = \mathrm{A} N(t) + \psi(t), \label{linea-qua}
\end{equation}
where $N(t)$ and $\psi(t)$ are the vectors of fluctuation operators and noise terms, respectively. Their explicit representations in terms of quadrature components are given as follows
\begin{align}
    N(t)=&\left[\delta x_a,\; \delta y_a,\;\delta x_m,\; \delta y_m, \;\delta q, \;\delta p\right]^T ,\\
    \psi(t)=& \left[\sqrt{2\kappa_a}\Tilde{x}_a, \sqrt{2\kappa_a}\Tilde{y}_a,\sqrt{2\kappa_m} x_m, \sqrt{2\kappa_m} y_m, 0, \zeta\right]^T ,
\end{align}
with the quadrature fluctuation operators are defined as $\delta x_a=(\delta a+\delta a^+)/\sqrt{2}$, $\delta x_a=(\delta a-\delta a^+)/i\sqrt{2}$, $\delta x_m=(\delta m+\delta m^+)/\sqrt{2}$ and $\delta x_m=(\delta m-\delta m^+)/i\sqrt{2}$, and the noise quadratures are given by $\Tilde{x}_a=(\Tilde{\eta}_a+\Tilde{\eta}_a^+)/\sqrt{2}$, $\Tilde{y}_a=(\Tilde{\eta}_a-\Tilde{\eta}_a^+)/i\sqrt{2}$, $x_m=(\eta_m+\eta_m^+)/\sqrt{2}$, and $y_m=(\eta_m-\eta_m^+)/i\sqrt{2}$.
The drift matrix $\mathrm{A}$ appearing in Eq. \eqref{linea-qua} is expressed as follows
\begin{equation}
	\mathrm{A}=\left(\begin{array}{cccccc}
               -\kappa_{a,fb} & \Delta_{a,fb} & 0 & g_{ma} & 0 & 0  \\
			 -\Delta_{a,fb} & -\kappa_{a,fb} & -g_{ma} & 0 & 0 & 0 \\
			 0 & g_{ma} & -\kappa_m & \Tilde{\Delta}_m & -g_{md} & 0  \\
			 -g_{ma} & 0 & -\Tilde{\Delta}_m & -\kappa_m & 0 & 0 \\
             0 & 0 & 0 & 0 & 0 & \omega_d \\
             0 & 0 & 0 & g_{md} & -\omega_d & -\gamma_d 
             
	\end{array}\right). \label{matrix A}
\end{equation}

Given the Gaussian nature of the input noise and the linearized dynamics, the steady state of the cavity–magnon–mechanical hybrid system is fully described by a three-mode Gaussian state. This state is completely characterized by a $6\times6$ covariance matrix  $\mathrm{V}$, whose elements are defined as 
\begin{equation}
    \mathrm{V}_{ij} = \langle N_i(\textit{t}), N_j(\textit{s}) + N_j(\textit{s}), N_i(\textit{t}) \rangle \hspace{0.5cm} i,j = 1,\dots,6
\end{equation}
Under the condition $|\Tilde{\Delta}_m|,\,|\Delta_{a,fb}|\gg \kappa_{a,fb},\,\kappa_m$, the steady-state covariance matrix is obtained by solving the following Lyapunov algebraic equation
\begin{equation}
    \mathrm{A} \mathrm{V} + \mathrm{V} \mathrm{A}^T=-\mathrm{D},\label{LPV}
\end{equation}
where the diffusion matrix $\mathrm{D}$ is defined by $\langle \psi_i(\textit{t})\,\psi_j(\textit{s})+\psi_j(\textit{s})\,\psi_i(\textit{t})\rangle/2= \mathrm{D}_{ij}\delta(\textit{t}-\textit{s})$. Using the noise correlations, we obtain
\begin{align}
    \mathrm{D}=&\,\text{diag} \big[\kappa_a \Lambda (2 \textit{n}(\omega_a)+1),\, \kappa_a \Lambda(2 \textit{n}(\omega_a)+1), \nonumber\\
    &\kappa_m(2 \textit{n}(\omega_m)+1),\kappa_m(2 \textit{n}(\omega_m)+1),\, 0,\, \gamma_s(2 \textit{n}(\omega_d)+1)\big],
\end{align}
where $\Lambda=\varepsilon^2 |1-r\,e^{i\theta}|^2$. In general, obtaining an analytical solution to the Lyapunov equation \eqref{LPV} is not feasible due to its complexity. Therefore, it is typically solved numerically to determine the steady-state covariance matrix, which is real, symmetric, and positive semidefinite. Its general structure can be written as
\begin{equation}
	\mathrm{V}=\left(\begin{array}{ccc}
             L_a  & \mathrm{V}_{am} & \mathrm{V}_{ad}  \\
             \mathrm{V}_{am}^T & L_m & \mathrm{V}_{md}  \\
             \mathrm{V}_{ad}^T & \mathrm{V}_{md}^T & L_d  
	\end{array}\right). \label{matrix V}
\end{equation}
The blocks $L_k \;(k=a,m,d)$ describe the individual dynamical features of each mode and have a dimension of $2\times2$. The off-diagonal blocks $ \mathrm{V}_{k,l}\;(k,l = a,m,d$ with $k \ne L)$ contain instead the information about the correlations established between the different modes, and are likewise $2\times2$ matrices. To guarantee that the system evolves towards a physically meaningful steady state, we impose stability conditions based on the Routh–Hurwitz criterion \cite{DeJesus1987}. In practice, this requirement ensures that the real parts of all eigenvalues of the drift matrix $\mathrm{A}$ remain negative, so that no parametric-feedback-induced instabilities can arise. Throughout the Results and Discussion section, only parameter configurations that fulfill these stability requirements are considered.

\section{APPLICATION}\label{APPLICATION}
In this section, we apply the theoretical formalism developed previously to our hybrid magnomechanical system, in which a coherent feedback loop is introduced to modify and optimize the overall dynamics of the device. Our main objective is to investigate how this feedback influences the precision of the simultaneous estimation of the two fundamental couplings of the system: the magnon–cavity coupling $g_{ma}$, which governs the coherent transfer of excitations between the magnon mode of the YIG crystal and the microwave field of the cavity, and the effective magnomechanical coupling $g_{md}$, arising from magnetostrictive deformation responsible for the interaction between magnons and mechanical vibrations. These parameters play a crucial role in the emergence of strong-coupling regimes, in the transfer of quantum information between subsystems, and in the generation of hybrid correlations. Their precise estimation is essential for properly characterizing the operation of the device and accessing the quantum regimes of interest.\\
A second objective is to determine which of the two metrological approaches—based on the RLD or SLD operators—provides the highest precision for the joint estimation of $g_{ma}$ and $g_{md}$. To assess the performance of the system, we compute both the CFIM and the QFIM associated with these two parameters. The corresponding CRB and QCRB allow us to identify the fundamental precision limits achievable in our configuration. The comparative analysis of these bounds highlights the role of coherent feedback: depending on the parameter regime, it can enhance or reduce the sensitivity of the system, thereby revealing the optimal conditions for robust multiparameter estimation. Based on Eqs. \eqref{FR with inv}, \eqref{FS with inv} and \eqref{CF}, we obtain the analytical expressions of the QFIM and CFIM used in our study, given as follows
\begin{equation}
\mathrm{F}^R =
 \begin{bmatrix}
       \mathrm{F}^R_{g_{ma},g_{ma}} &    
       \mathrm{F}^R_{g_{ma},g_{md}} \\
     \\
       \mathrm{F}^R_{g_{md},g_{ma}} &
       \mathrm{F}^R_{g_{md},g_{md}} 
\end{bmatrix}, \label{QFIM-RLD fin}
\end{equation}
\begin{equation}
\mathrm{F}^S =
 \begin{bmatrix}
       \mathrm{F}^S_{g_{ma},g_{ma}} &    
       \mathrm{F}^S_{g_{ma},g_{md}} \\
     \\
       \mathrm{F}^S_{g_{md},g_{ma}} &
       \mathrm{F}^S_{g_{md},g_{md}} 
\end{bmatrix},\label{QFIM-SLD fin}
\end{equation}
and
\begin{widetext}
    \begin{equation}
       \mathrm{F}^C = \frac{1}{2}
       \begin{bmatrix}
      \operatorname{Tr}\left[ \mathcal{V}^{-1} \frac{\partial \mathcal{V}}{\partial g_{ma}} \mathcal{V}^{-1} \frac{\partial \mathcal{V}}{\partial g_{ma}} \right] + \left( \frac{\partial \langle R \rangle}{\partial g_{ma}} \right)^T \mathcal{V}^{-1} \frac{\partial \langle R \rangle}{\partial g_{ma}} &\hspace{0.2cm}    
      \operatorname{Tr}\left[ \mathcal{V}^{-1} \frac{\partial \mathcal{V}}{\partial g_{ma}} \mathcal{V}^{-1} \frac{\partial \mathcal{V}}{\partial g_{md}} \right] + \left( \frac{\partial \langle R \rangle}{\partial g_{ma}} \right)^T \mathcal{V}^{-1} \frac{\partial \langle R \rangle}{\partial g_{md}} \vspace{0.5cm}
     \\ 
      \operatorname{Tr}\left[ \mathcal{V}^{-1} \frac{\partial \mathcal{V}}{\partial g_{md}} \mathcal{V}^{-1} \frac{\partial \mathcal{V}}{\partial g_{ma}} \right] + \left( \frac{\partial \langle R \rangle}{\partial g_{md}} \right)^T \mathcal{V}^{-1} \frac{\partial \langle R \rangle}{\partial g_{ma}} &\hspace{0.2cm}
      \operatorname{Tr}\left[ \mathcal{V}^{-1} \frac{\partial \mathcal{V}}{\partial g_{md}} \mathcal{V}^{-1} \frac{\partial \mathcal{V}}{\partial g_{md}} \right] + \left( \frac{\partial \langle R \rangle}{\partial g_{md}} \right)^T \mathcal{V}^{-1} \frac{\partial \langle R \rangle}{\partial g_{md}} 
\end{bmatrix}
\end{equation}
The elements of the QFIMs, $\mathrm{F}^R$ and $\mathrm{F}^S$, given in Eqs. \eqref{QFIM-RLD fin} and \eqref{QFIM-SLD fin}, are calculated as follows
\begin{align}
     \mathrm{F}_{g_{ma} g_{md}}^R =& \;2\, \text{Vec}\left[\frac{\partial \mathrm{V}}{\partial{g_{ma}}} \right]^\dagger \left((2\mathrm{V}+i\Omega)^\dagger\otimes(2\mathrm{V}+i\Omega)\right)^{-1} \text{Vec}\left[\frac{\partial \mathrm{V}}{\partial{g_{md}}} \right]+ 2 \left(\frac{\partial \langle R \rangle}{\partial{g_{ma}}}\right)^T   (2\mathrm{V}+i\Omega)^{-1} \left(\frac{\partial \langle R \rangle}{\partial{g_{md}}}\right),\\
      \mathrm{F}_{g_{ma} g_{md}}^S =& \, 2\,\text{Vec}\left[\frac{\partial \mathrm{V}}{\partial{g_{ma}}} \right]^\dagger  
     \left(4\mathrm{V}^\dagger \otimes \mathrm{V} + \Omega \otimes \Omega \right)^{-1}  \text{Vec}\left[\frac{\partial \mathrm{V}}{\partial{g_{md}}} \right] 
     +  \left(\frac{\partial \langle R \rangle}{\partial{g_{ma}}}\right)^T  \mathrm{V}^{-1}\left(\frac{\partial \langle R \rangle}{\partial{g_{md}}}\right).
\end{align}
\end{widetext}
In the following, we perform our calculations fully numerically, since the matrix structures involved in the expressions of the CFIM and QFIM are too complex to allow reliable analytical inversion. To identify which estimation strategy provides the highest precision, we analyze the ratio $\mathrm{C}^R/\mathrm{C}^S$. The study is carried out using a set of experimentally feasible parameters \cite{Zhang2016,Li2018,Zhang2014,Zuo2024,Rameshti2022,Rondin2014}: the cavity and magnon frequencies are set to $\omega_a/2\pi=\omega_m/2\pi=10 \,\text{GHz}$, while the mechanical frequency is chosen as $\omega_d/2\pi=10 \,\text{MHz}$. The corresponding dissipation rates are $\kappa_a/2\pi=\kappa_m/2\pi=0.1\;\text{MHz}$ and $\gamma_d/2\pi=10\;\text{KHz}$. The two coupling strengths are initially taken equal, $g_{ma}/2\pi=g_{md}/2\pi=2\;\text{MHz}$. Finally, the cavity and magnon detunings are chosen proportional to the mechanical resonance frequency, such that $\Delta_a = 0.9\,\Delta_m = \omega_d$. The feedback parameters are varied within the following ranges: the reflection coefficient $r\in\left[0,\, 0.5\right]$ and the phase shift $\theta\in\left[0,\, 2\pi\right]$. This configuration serves as the basis for a quantitative comparison between the RLD- and SLD-based approaches in the simultaneous estimation of the two couplings of the system.\par

\begin{figure}[h]
    \centering

    \includegraphics[width=0.9\linewidth]{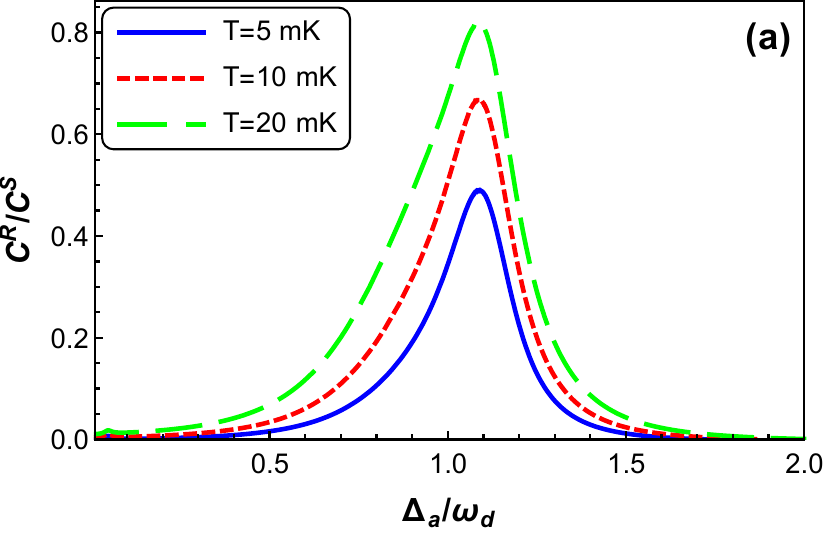}
    
    \includegraphics[width=0.9\linewidth]{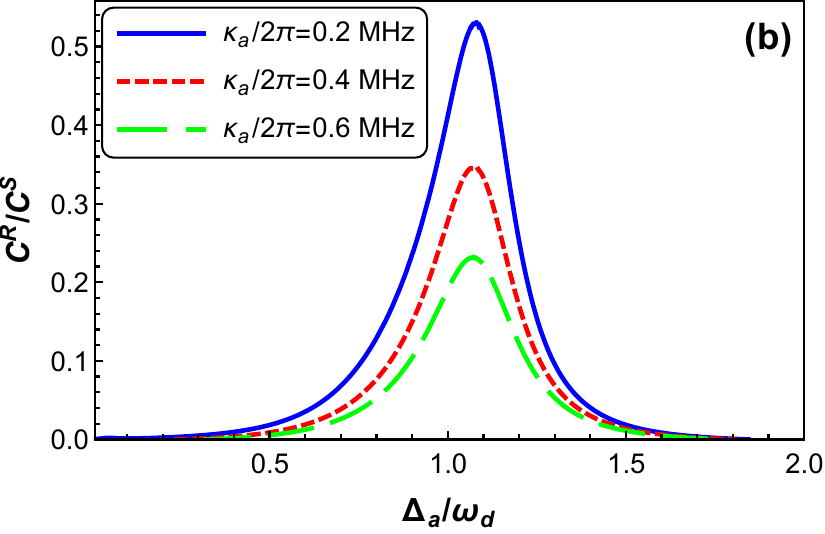}

    \caption{Ratio between the RLD-QCRB and SLD-QCRB as a function of the normalized cavity detuning $\Delta_a$. Panel (a) illustrates the dependence of this ratio on the environmental temperature $T$, while panel (b) presents its variation with the cavity dissipation rate $\kappa_a$. The pump power is fixed at $P=8.9 \;\text{mW}$, and the feedback parameters are set to $r=0.1$ and $\theta=\pi$. In panel (b), the temperature is kept constant at $T = 10 \,\text{mK}$.}\label{rati}
\end{figure}

Fig. \eqref{rati} displays the behavior of the ratio between the RLD-QCRB and the SLD-QCRB as a function of the normalized cavity detuning $\Delta_a$. Panel (a) examines this ratio for different temperatures $T$, while panel (b) shows its dependence on various cavity dissipation rates $\kappa_a$. In both cases, the ratio $\mathrm{C}^R/\mathrm{C}^S$ increases with temperature and decreases as  $\kappa_a$ is reduced. It reaches its maximum around $\Delta_a \approx 1.1\omega_d$, then drops as $\Delta_a$ moves further away from this point, and eventually vanishes for $\Delta_a\leq0$ and $\Delta_a\ge2 \omega_d$. These results reveal that, for all considered values of the detuning $\Delta_a$, the temperature $T$, the dissipation rate $\kappa_a$, and the remaining system parameters used in this study, the ratio remains strictly below unity. This means that the most informative bound is systematically the RLD-QCRB $(\mathrm{C}^R)$, which coincides with the $\mathrm{C}^{MI}$ bound in our configuration. We have also verified numerically that this conclusion remains valid across the entire parameter space explored. It should be emphasized, however, that this behavior is not universal: in other physical platforms, the ratio $\mathrm{C}^R/\mathrm{C}^S$ may become greater than or equal to one, in which case the SLD-QCRB would be the tighter bound. Overall, our results demonstrate that, in the cavity–magnon–mechanical system considered here, the RLD-based quantum Cramér–Rao bound consistently provides superior estimation performance compared to the SLD-based bound. From a physical perspective, the ratio $\mathrm{C}^R/\mathrm{C}^S$ quantifies the relative efficiency of the two quantum bounds, offering a direct tool to identify the optimal estimation strategy under varying operational conditions. This comparative framework is therefore essential for selecting the appropriate metrological approach and enhancing estimation precision in future experimental implementations.

\begin{figure}[h]
    \centering

    \includegraphics[width=1\linewidth]{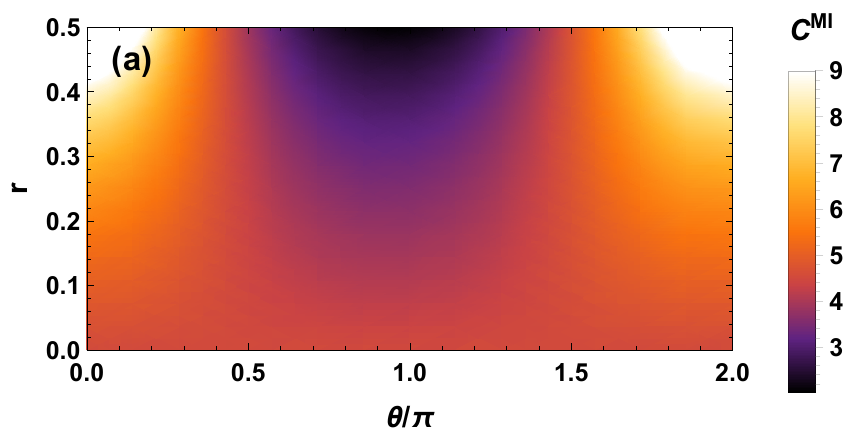}
    
    \includegraphics[width=1\linewidth]{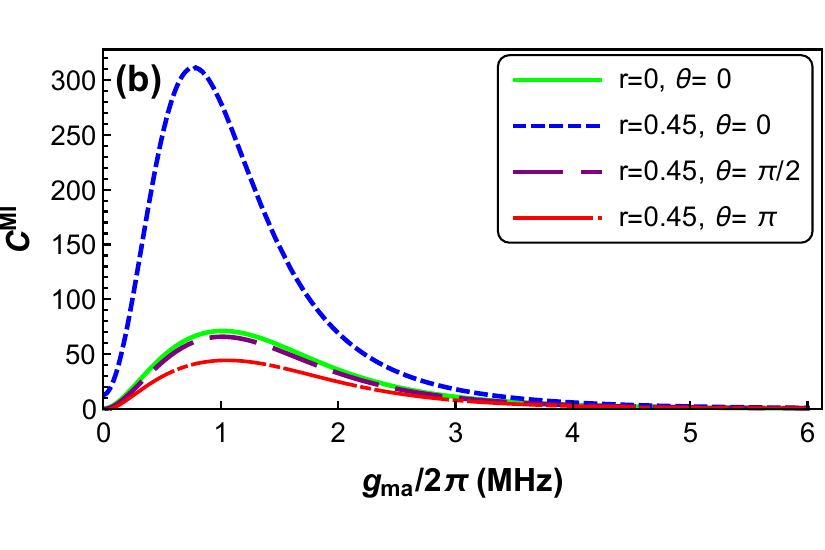}
    
    \includegraphics[width=1\linewidth]{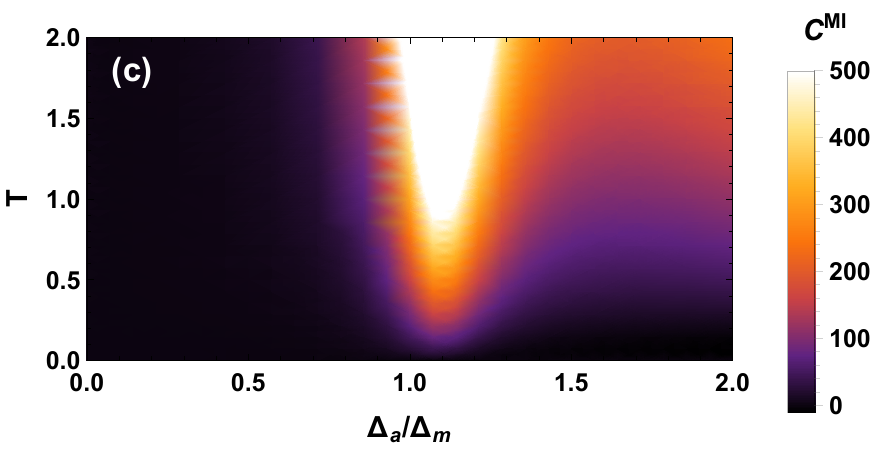}

    \caption{(a) Density plot of the $\mathrm{C}^{MI}$ as a function of the feedback parameters: the reflection coefficient $r$ and the phase shift $\theta$. (b) $\mathrm{C}^{MI}$ as a function of the magnon–cavity coupling strength $g_{ma}$ for different feedback configurations. (c) Density plot of the $\mathrm{C}^{MI}$ as a function of the detuning ratio $\Delta_a/\Delta_m$ and temperature $T$ (in K), with $r=0.5$ and $\theta=\pi$. In panels (a) and (b), the temperature is fixed at $T=10$ mK. All other system parameters are the same as in Fig. \ref{rati}.}
    \label{feedback}
\end{figure}

In the second part of our analysis, we demonstrate that the precision in estimating the coupling parameters $g_{ma}$ and $g_{md}$ can be significantly enhanced by introducing coherent feedback into the cavity–magnon–mechanical system. Fig. \ref{feedback}(a) and \ref{feedback}(b) show the dependence of the $\mathrm{C}^{MI}$ (which corresponds to the most informative QCRB) on the reflection coefficient $r$ and the feedback phase $\theta$. To ensure the dynamical stability of the system, the range of $r$ is restricted to $0\leq r\leq0.5$. For $r>0.5$, the effective cavity dissipation rate $\kappa_{a,fb}$ may become negative for certain phases, leading to instability. Fig. \ref{feedback}(a) reveals that the estimation precision of both coupling strengths can be substantially improved when the feedback parameters are optimally chosen. Conversely, an inappropriate choice can significantly deteriorate the performance. For instance, the $\mathrm{C}^{MI}$ reaches its minimum (corresponding to minimal estimation error) at $r=0.5$ and $\theta=\pi$, representing a marked improvement compared to the no-feedback case $(r=0)$. In contrast, for $r=0.5$ and $\theta=0$ or $2\pi$, the $\mathrm{C}^{MI}$ increases, indicating poorer precision. According to Fig. \ref{feedback}(a), the most beneficial parameter region typically lies within $r\in [0.3, 0.5]$ and $\theta/2\pi \in [0.5, 1.5]$. The optimal phase is $\theta=\pi+2n\pi\;(n\in \mathbb{Z})$, as it minimizes the diffusion of the cavity mode, thereby strengthening quantum coherence and improving the estimation accuracy. However, this phase does not necessarily reduce $\kappa_{a,fb}$, and a reduced dissipation rate does not always correlate with better performance. Conversely, $\theta=2n\pi\;(n\in \mathbb{Z})$ maximizes diffusion, increasing the estimation error and significantly degrading precision for both couplings.\\
Fig. \ref{feedback}(b) shows that the $\mathrm{C}^{MI}$ exhibits a maximum at $g_{ma} /2\pi = 1\;\text{MHz}$, indicating large estimation error at this specific point. Beyond this value, the $\mathrm{C}^{MI}$ decreases, which corresponds to an enhancement of the Fisher information and thus an improvement in parameter estimation. Comparison between the curves highlights that adding coherent feedback leads to a substantial reduction in the $\mathrm{C}^{MI}$, confirming a clear enhancement in the precision of estimating $g_{ma}$ and $g_{md}$.\\
Finally, Fig. \ref{feedback}(c) shows that the  $\mathrm{C}^{MI}$ can reach very large values—up to $\mathrm{C}^{MI}\approx 500$ at a temperature of $T=2K$—when the detuning condition $\Delta_a=1.1\,\Delta_m$ is satisfied. This behavior persists with or without coherent feedback, suggesting that the optimal detuning condition is an intrinsic property of the magnomechanical system. This invariance arises because, under optimal feedback $(r=0.5,\; \theta=\pi)$, the effective detuning remains unchanged, $\Delta_{a,fb}=\Delta_a$, thus preserving a resonance condition favorable for information transfer among the magnon, cavity, and mechanical modes. For $0\leq\Delta_a/\Delta_m\leq 0.8$, the $\mathrm{C}^{MI}$ is small, indicating that the estimation error is nearly negligible and that the coupling estimation is highly precise. In this regime, hybridization between modes is limited, and the system becomes weakly sensitive to variations in the coupling strengths. Conversely, for $1.5\leq\Delta_a/\Delta_m\leq 2$ and at low temperatures $T<0,5 \,\text{K}$, the Fisher information increases significantly, leading to a more accurate estimation of the parameters. The enhanced hybridization and the reduced effective dissipation induced by coherent feedback reinforce the system’s coherence, thereby improving estimation precision even under thermal noise. Overall, these results demonstrate that coherent feedback not only amplifies the $\mathrm{C}^{MI}$ but also strengthens its robustness against thermal fluctuations, making the proposed scheme highly promising for the precise estimation of magnomechanical coupling parameters.

\begin{figure}[H]
    \centering

    \includegraphics[width=0.9\linewidth]{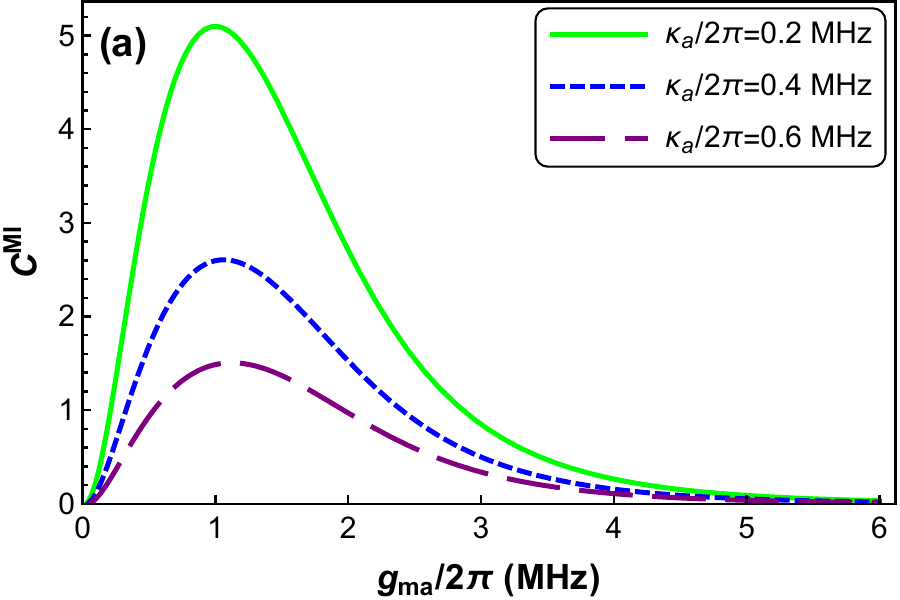}
    
    \includegraphics[width=0.9\linewidth]{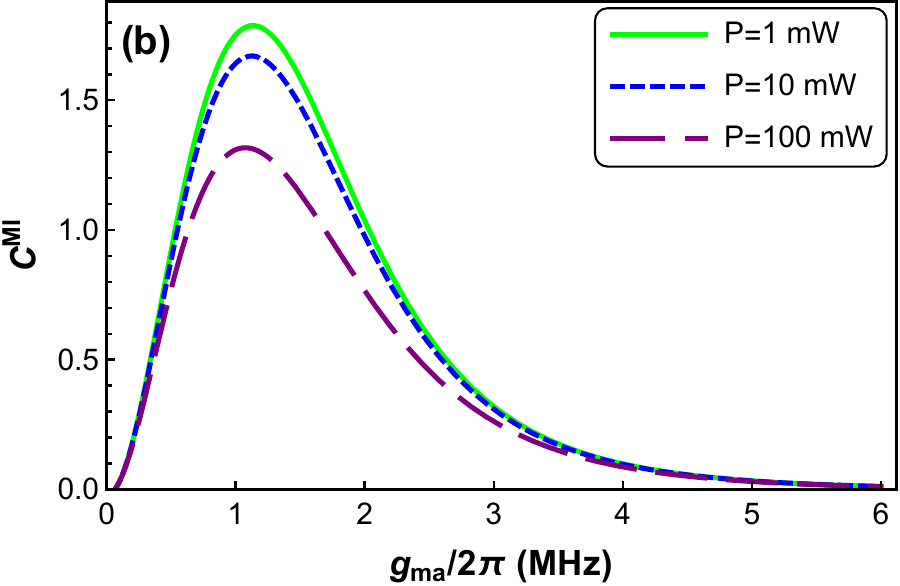}

    \caption{The most informative-QCRB $\mathrm{C}^{MI}$ as a function of the magnon–cavity coupling strength $g_{ma}$: (a) for different values of the cavity decay rate $\kappa_a$, and (b) for different input powers $P$, with $T = 10 \,\text{mK}$. In panel (b), we set $\kappa_a/2\pi=\kappa_m/2\pi= 0.4 \,\text{MHz}$, while all other parameters are the same as in Fig. \ref{rati}.}\label{CMI-gma}
\end{figure}
The results presented in Fig. \ref{CMI-gma} reveal how the input driving power $P$ and the cavity decay rate $\kappa_a$ influence the precision of estimating the magnon–cavity coupling $g_{ma}$ and the magnomechanical coupling $g_{md}$ in a cavity–magnomechanical platform operating under coherent feedback. To analyze this behavior, we examine the evolution of the $\mathrm{C}^{MI}$ as a function of the magnon–cavity interaction strength $g_{ma}$. In the region  $0<g_{ma}/2\pi\leq 1 \,\text{MHz}$, the $\mathrm{C}^{MI}$ increases with $g_{ma}$. This trend reflects the fact that a moderate enhancement of the photon–magnon hybridization improves the ability of the system to encode information about the parameters to be estimated. At $g_{ma}/2\pi\approx 1 \,\text{MHz} $, the $\mathrm{C}^{MI}$ reaches its maximum, marking a regime where the system exhibits minimal sensitivity to variations in $g_{ma}$ and $g_{md}$. Consequently, the multiparameter estimation error becomes largest, indicating that only a limited amount of Fisher information is accessible at this particular coupling strength. For stronger interactions, $g_{ma}/2\pi> 1 \,\text{MHz} $, the $\mathrm{C}^{MI}$ decreases rapidly. In this domain, the system enters a strong-coupling regime in which the hybridization between the magnon and cavity modes becomes dominant, effectively rigidifying the system dynamics. This reduced dynamical sensitivity leads to a significant suppression of the estimation error—$\mathrm{C}^{MI}$ approaches zero—corresponding to a substantial increase in the available Fisher information. In this regime, the system becomes highly informative about the parameters $g_{ma}$ and $g_{md}$. 
Furthermore, increasing either the cavity decay rate $\kappa_a$ or the input power $P$ results in a systematic reduction of the $\mathrm{C}^{MI}$. Physically, this behavior indicates an enhancement of the information encoded in the output field. A stronger drive amplifies both magnon–photon and magnomechanical interactions, while appropriate dissipation can stabilize the Feedback-modified dynamics and enhance the relative contribution of useful quantum fluctuations over noise. As a result, larger values of $P$ and suitable values of $\kappa_a$ improve the estimation precision by lowering the error associated with both coupling parameters. 
Overall, these observations highlight how the interplay between coupling strengths, dissipation, and coherent feedback determines the optimal operating regime for high-precision multiparameter estimation in cavity–magnomechanical systems.

\begin{figure}[h]
    \centering

    \includegraphics[width=0.9\linewidth]{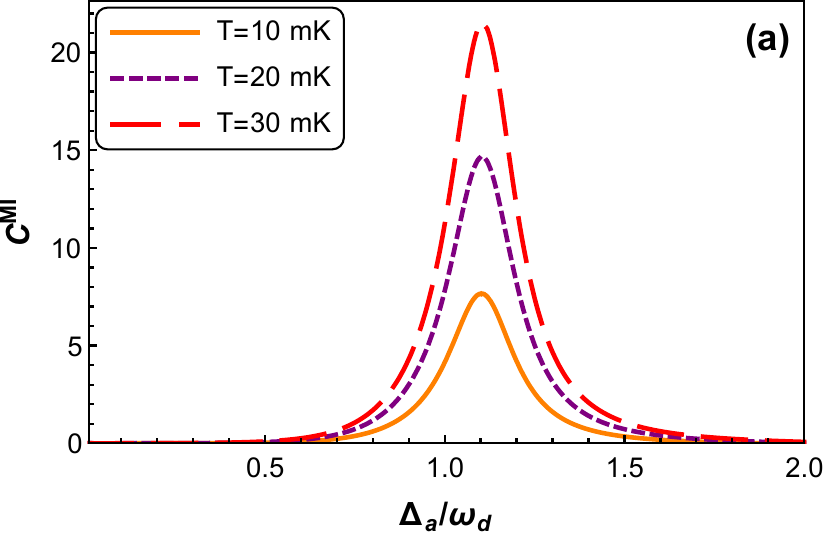}
    
    \includegraphics[width=0.9\linewidth]{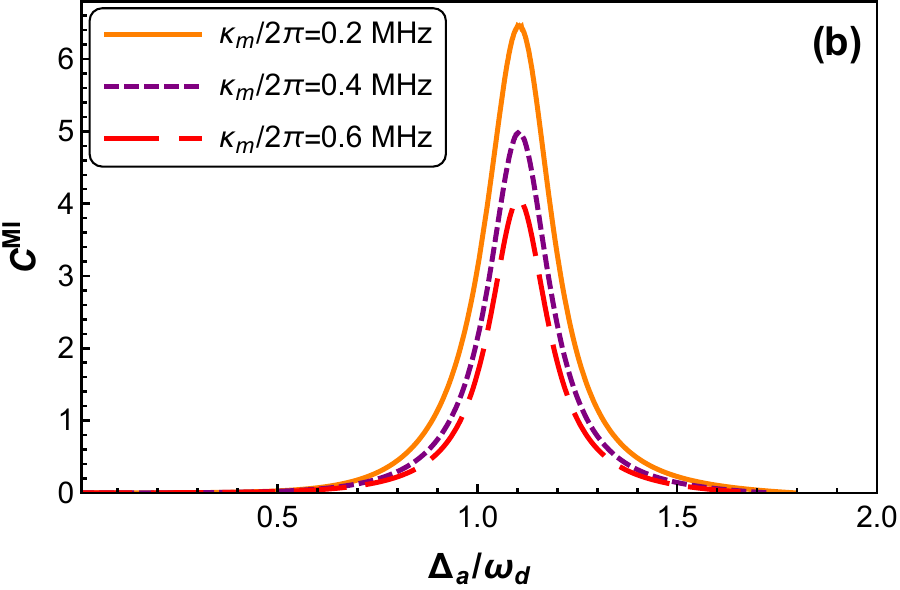}

    \caption{Plot of the most informative-QCRB $\mathrm{C}^{MI}$ as a function of the normalized cavity detuning, $\Delta_a$. Panel (a) shows its variation for different values of the environmental temperature $T$. Panel (b) shows its variation for different values of the mechanical mode decay rate $\kappa_m$. The other system parameters are identical to those used in Fig. \ref{rati}.}\label{2CMI-Da}
\end{figure}
Fig. \ref{2CMI-Da} provides a detailed analysis of how the $\mathrm{C}^{MI}$ responds to variations in temperature and mechanical dissipation in our cavity–magnon–mechanical system operating under coherent feedback. In these plots, the $\mathrm{C}^{MI}$ is shown as a function of the normalized detuning $\Delta_a/\omega_d$, for different values of the environmental temperature $T$ in panel (a), and for various mechanical loss rates $\kappa_m$ in panel (b). Both panels reveal the emergence of a pronounced $\mathrm{C}^{MI}$ peak around $\Delta_a\approx 1.1\,\omega_d$ for fixed values of $T$ and $\kappa_m$. This peak reflects a significant increase in the $\mathrm{C}^{MI}$, corresponding to a larger estimation error for the coupling parameters $g_{ma}$ and $g_{md}$. In this region, the system becomes less sensitive to variations in the coupling strengths, thereby degrading the precision of the estimation. When the detuning is fixed at $\Delta_a = 1.1\,\omega_d$, the $\mathrm{C}^{MI}$ is observed to decrease as the temperature $T$ is lowered and as the mechanical dissipation rate $\kappa_m$ is increased. A lower $\mathrm{C}^{MI}$ implies a reduced estimation error, which translates into a notable improvement in the precision with which the magnomechanical couplings can be estimated. This behavior arises because, at low temperatures, thermal occupations become negligible and quantum correlations between the hybrid modes are better preserved. At the same time, a controlled increase in mechanical dissipation can suppress certain parasitic fluctuations, thereby reinforcing the coherence among the magnonic, mechanical, and cavity excitations. In the intervals $0\leq\Delta_a/\omega_d\leq 1.5$ and $1.7\leq\Delta_a/\omega_d\leq 2$, the $\mathrm{C}^{MI}$ is very small, and in some cases nearly zero. This corresponds to a high QFI for the couplings $g_{ma}$ and $g_{md}$, indicating that parameter estimation is particularly precise within these regions. The low $\mathrm{C}^{MI}$ in these domains reflects an enhanced flow of information between the system modes, facilitated by coherent feedback, which reduces the impact of thermal and dissipative effects and improves the overall sensitivity of the estimation protocol.

\begin{figure}[h]
    \centering

    \includegraphics[width=0.9\linewidth]{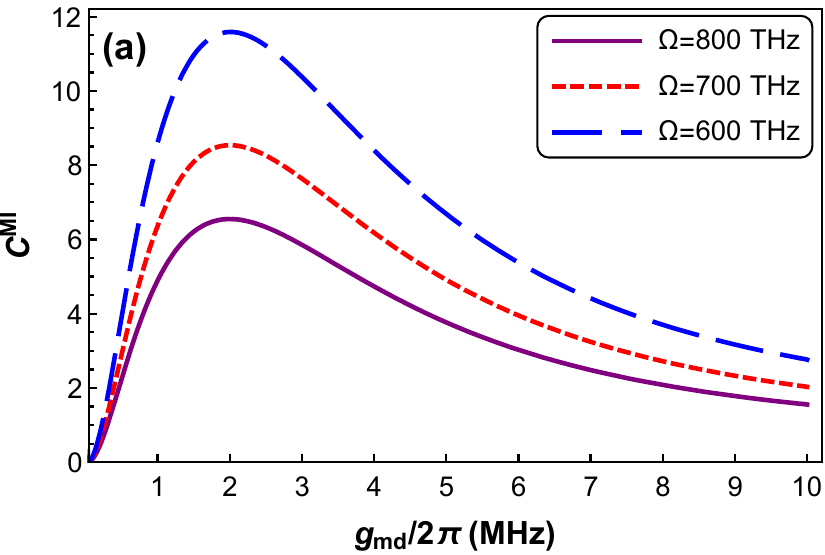}
    
    \includegraphics[width=0.9\linewidth]{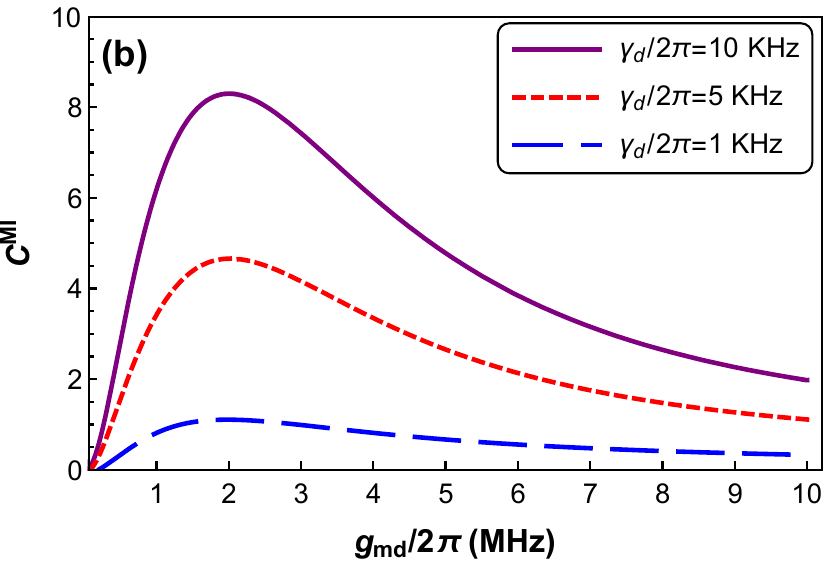}

    \caption{$\mathrm{C}^{MI}$ as a function of the effective magnomechanical coupling strength $g_{md}$. Panel (a) shows the dependence on the Rabi frequency $\Omega$, while panel (b) displays the variation with the mechanical damping rate $\gamma_d$. The environmental temperature is fixed at $T=10 \;\text{mK}$, and the magnon-cavity coupling is set to $g_{ma}/2\pi=1\;\text{MHz}$. All remaining parameters are identical to those in Fig. \ref{rati}.}\label{2CMI-gd}
\end{figure}
Here, we focus on the influence of the Rabi frequency $\Omega$ and the mechanical damping rate $\gamma_d$ on the estimation precision of the magnon–cavity coupling $g_{ma}$ and the magnon–mechanical coupling $g_{md}$. Fig. \ref{2CMI-gd} presents the most informative-QCRB for the estimation of these two couplings as a function of the effective magnon–mechanical interaction strength, for different values of the Rabi frequency and the mechanical damping rate. From these results, we observe that when $\Omega$ and $\gamma_d$ are kept fixed, the quantity $\mathrm{C}^{MI}$ reaches a maximum around the point $g_{md}\approx2\pi\times2\,\text{MHz}$. A larger value of $\mathrm{C}^{MI}$ indicates a larger estimation error, which means that the Fisher information about the parameters is strongly suppressed in this region. Beyond this point, $\mathrm{C}^{MI}$ decreases as $g_{md}$ increases, which corresponds to an enhancement of the estimation precision for both $g_{ma}$ and $g_{md}$.\\
When examining the roles of $\Omega$ and $\gamma_d$, we find that $\mathrm{C}^{MI}$ decreases with increasing Rabi frequency and decreasing mechanical damping rate. Physically, a larger $\Omega$ strengthens the interaction between the cavity field and the magnon mode, enhancing coherence and increasing the amount of useful quantum information imprinted on the output field. Similarly, reducing $\gamma_d$ suppresses mechanical dissipation, which preserves the correlations shared between the mechanical mode and the magnonic system. Both effects result in an increase of the Fisher information and therefore lead to a significant improvement of the metrological precision in estimating the coupling parameters.


\begin{widetext}
\begin{figure*}[t]
\centering
\includegraphics[width=0.24\textwidth]{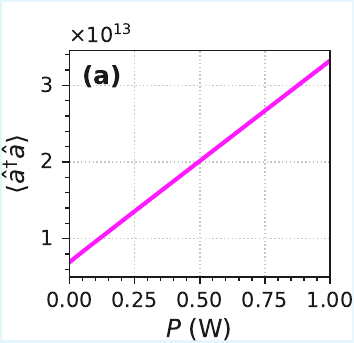}\hfill
\includegraphics[width=0.255\textwidth]{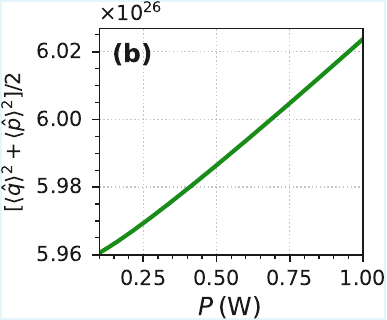}\hfill
\includegraphics[width=0.245\textwidth]{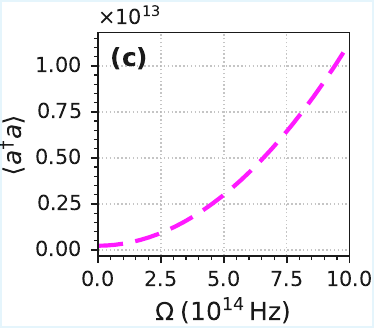}\hfill
\includegraphics[width=0.24\textwidth]{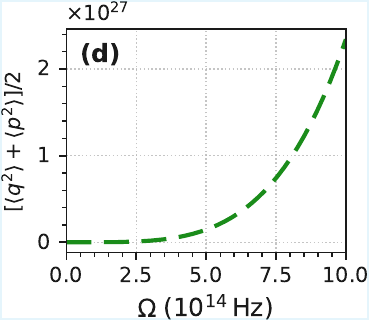}
\caption{Average number of photons (a, c) and phonons (b, d) as functions of (a, b) the input power $P$ and (c, d) the Rabi frequency $\Omega$. The parameters used are the same as those in Fig. \ref{rati}.}
\label{photon and phonon}
\end{figure*}

\end{widetext}

To further confirm the previously discussed results and clarify the physical origin of the observed enhancement in estimation precision, we analyze the impact of the optical power $P$ and the Rabi frequency $\Omega$ on the magnonic-cavity system. Figs. \ref{photon and phonon} clearly show that both parameters significantly influence the estimation precision in the cavity-magnomechanical system. Increasing the optical power $P$ leads to a rise in the average number of photons and phonons within the system. Due to the intrinsic coupling between photons and magnons, this increase directly results in a higher average number of magnons. Consequently, the enhanced populations strengthen both the magnon-cavity and magnon-mechanical interactions, thereby increasing the system’s capacity to store quantum information. This amplification of the relevant mode populations makes more information available for estimating the coupling parameters $g_{ma}$ and $g_{md}$, as illustrated in Fig. \ref{CMI-gma} (b), which in turn improves the overall estimation precision. Similarly, the increase in the Rabi frequency $\Omega$ acts as a key mechanism through which the driving field enhances parameter estimation. A higher $\Omega$ increases the average photon and phonon populations, which simultaneously boosts magnon populations. This strengthens the optomagnon and magnon-mechanical interactions, effectively increasing the couplings $g_{ma}$ and $g_{md}$. As a result, the QFI for both parameters rises, while the QCRB decreases, as shown in Fig. \ref{2CMI-gd} (a). Indeed, a higher number of quanta allows for better state discrimination and more efficient extraction of quantum information, thus enhancing the system’s ability to estimate its parameters accurately.\\
Therefore, these results support the overall objective of our study —optimizing the precision in estimating the optomagnon coupling $g_{ma}$ and the magnon-mechanical coupling $g_{md}$— by highlighting how specific physical parameters, such as the optical power and Rabi frequency, influence the information carriers governing estimation efficiency. The positive correlation between estimation precision and the parameters $P$ and $\Omega$ is further corroborated by the results presented in Figs. \ref{CMI-gma} (b) and \ref{2CMI-gd} (a).

\begin{figure}[H]
    \centering

    \includegraphics[width=0.9\linewidth]{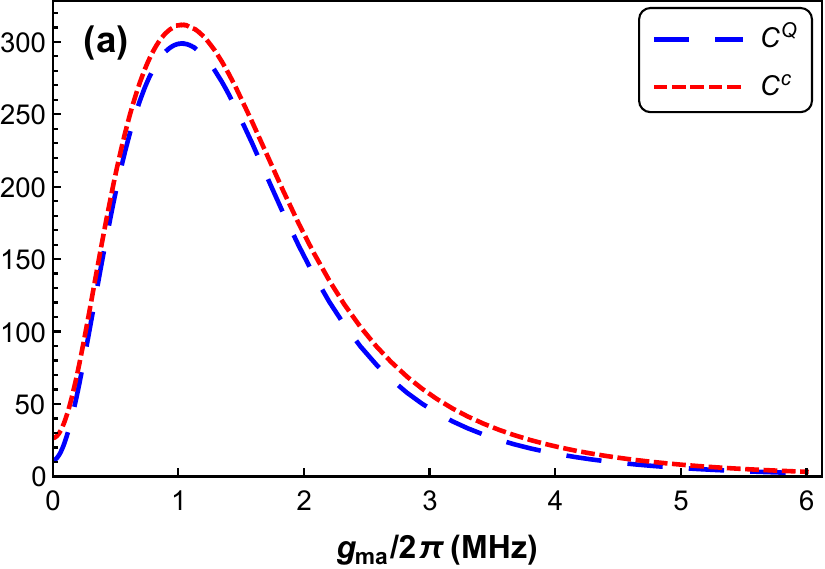}
    
    \includegraphics[width=0.9\linewidth]{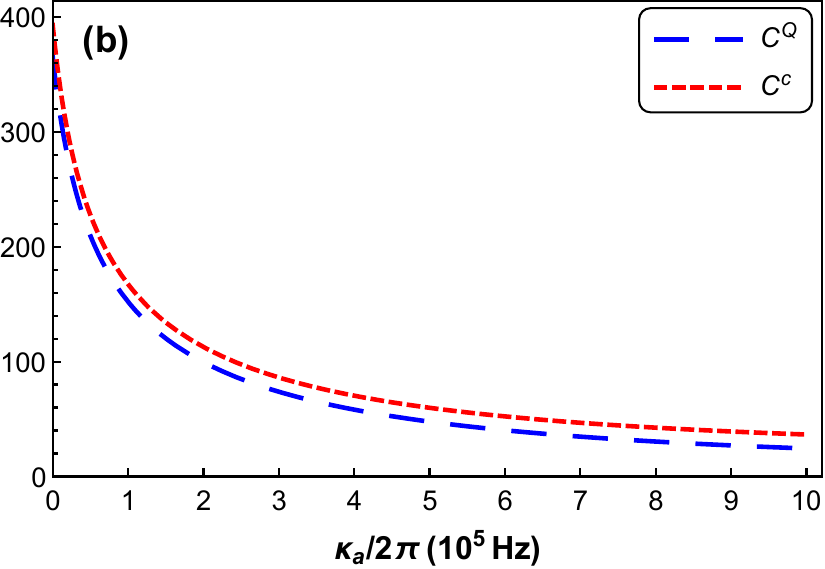}

    \caption{ Quantum Cramér–Rao bound $(\mathrm{C}^Q)$ and classical Cramér–Rao bound $(\mathrm{C}^Q)$ obtained via heterodyne detection as functions of (a) the cavity–magnon coupling $g_{ma}$ and (b) the cavity damping rate $\kappa_a$, for the simultaneous estimation of the couplings $g_{ma}$ and $g_{md}$. The environmental temperature is fixed at $T=0.5\;\text{K}$, while all other parameters are the same as those used in Fig. \ref{rati}.}\label{CQC}
\end{figure}
To compare the quantum Cramér–Rao bound (QCRB) of our system with the classical Cramér–Rao bound (CRB) based on heterodyne detection, Fig. \ref{CQC} displays both bounds for the multiparameter estimation of the cavity–magnon coupling $g_{ma}$ and the magnon–mechanical coupling $g_{md}$. Fig. \ref{CQC}(a) shows the dependence of the bounds on the cavity–magnon coupling $g_{ma}$, while Fig. \ref{CQC}(b) illustrates their behavior as a function of the optical damping rate $\kappa
_a$. In both panels, the blue curve corresponds to the QCRB, whereas the red curve represents the CRB obtained from heterodyne detection.\\
As shown in Fig. \ref{CQC}(a), both bounds reach a maximum around $g_{ma}/2\pi\approx 1\;\text{MHz}$, indicating a peak in the estimation error for the two coupled parameters at this point. Beyond this value, the bounds decrease with increasing $g_{ma}$, reflecting a progressive improvement in estimation precision. This behavior can be attributed to the strengthening of the system interactions, which enhances the amount of information available about the parameters.\\
Similarly, Fig. \ref{CQC}(b) shows that for small values of the optical damping rate $\kappa_a$, both the classical and quantum bounds take relatively large values, and then decrease gradually as $\kappa_a$ increases. This reduction of the bounds is associated with an increase in the Fisher information, indicating that, in this regime, controlled dissipation facilitates information extraction and improves estimation accuracy.\\
It is also worth emphasizing that the CRB and QCRB curves remain very close over the entire parameter range considered. In particular, the CRB (red curve) is always greater than or equal to the QCRB (blue curve), confirming the validity of the general inequality $\mathrm{C}^Q\leq \mathrm{C}^C$. This close agreement demonstrates that heterodyne detection provides a nearly optimal measurement strategy in our system, enabling efficient simultaneous estimation of the couplings $g_{ma}$ and $g_{md}$.

\section{Experimental feasibility}\label{Experimental feasibility}

The experimental feasibility of the proposed nonreciprocal quantum metrology scheme relies on well-established advances in cavity–magnon systems. In particular, magnon frequency shifts induced by the Kerr effect have already been reported experimentally in configurations that are fully compatible with the present proposal \cite{Wang2016}. Moreover, several fundamental quantum effects in cavity–magnon platforms have been successfully demonstrated, including the magnon-spring effect and dynamical backaction \cite{Yuan2022}. These experimental achievements represent essential milestones toward the realization of nonreciprocal quantum metrology protocols.\par 
In our setup, the bias magnetic field can be generated using electromagnets, while its orientation can be precisely controlled by adjusting the electric current in a Helmholtz coil. It is worth emphasizing that all parameters employed in our numerical simulations fall within ranges accessible to state-of-the-art experimental techniques. More generally, the feasibility of the proposed hybrid cavity–magnonic scheme is strongly supported by recent experimental progress \cite{Pozar2021,Marpaung2013,Marpaung2019}. In particular, strong coherent coupling between magnons in yttrium iron garnet (YIG) spheres and microwave photons in high-quality-factor cavities has been robustly demonstrated \cite{Zhang2014,Tabuchi2014,Huebl2013}.
\begin{figure}[H]
    \centering
    \includegraphics[width=0.5\textwidth]{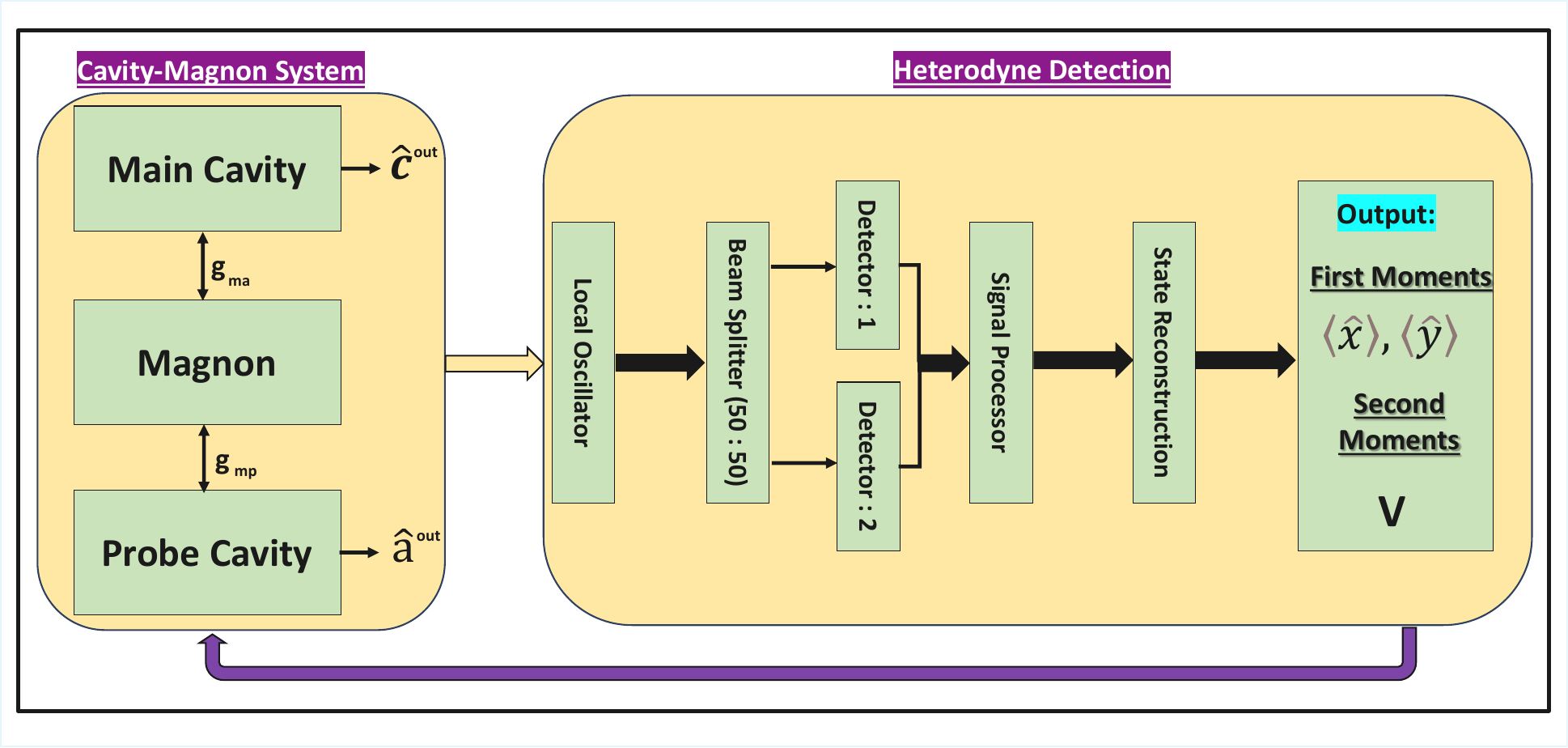}
    \caption{Schematic diagram illustrating the reconstruction of the first-order moments and the covariance matrix of a Gaussian state using heterodyne detection. Here, $g_{mp}$ denotes the coupling strength between the magnon mode and the probe cavity. The reconstruction procedure is based on the input–output relations \cite{Gardiner2004}.} \label{Hetero}
\end{figure}

In addition, coherent feedback control has already been successfully implemented in optomechanical platforms \cite{Ernzer2023,Du2025}. In close analogy, our coherent feedback loop is built from standard microwave photonic components—including beam splitters, highly reflective mirrors, and tunable phase shifters—which are well-established elements of contemporary microwave engineering \cite{Pozar2021,Sliwa2015,Gu2017,Barzanjeh2017}. These components enable a realistic and controllable implementation of the proposed feedback scheme.

Since the system under consideration is Gaussian, its quantum state is fully characterized by the first-order moments (quadratures) and the covariance matrix. These quantities can be accessed experimentally using homodyne or heterodyne detection, both of which are mature and widely employed Gaussian measurement techniques \cite{Serafini2023,Gardiner2004}. Direct measurement of the magnon mode is more challenging; however, it can be performed indirectly by coupling the YIG sphere to an auxiliary cavity driven by a weakly red-detuned laser. In this configuration, information about the magnonic degrees of freedom is encoded in the output field of the readout cavity. During the measurement process, the output quadratures of each cavity field exhibit stationary temporal traces, from which the stationary probability distributions—and hence the first-order moments—can be reconstructed \cite{Aspelmeyer2014}.

The covariance matrix is obtained by statistically analyzing the temporal traces, multiplying different quadratures and averaging their products. This procedure is implemented using the input–output formalism together with appropriate spectral filtering \cite{Aspelmeyer2014}. The resulting first-order moments and covariance matrix provide all the experimental data required to extract the quantum Fisher information (QFI) and to evaluate the ultimate precision bounds of the metrological protocol.

In our analysis, the optimal operating regime corresponds to a coherent feedback reflectivity $r=0.5$ and a feedback phase $\theta=\pi$, while the phase of the driving electromagnetic field is set to $\phi=0$. These parameter values are fully compatible with current experimental capabilities. Indeed, hybrid microwave beam splitters and couplers can achieve reflection coefficients close to unity, making the chosen value $r=0.5$ experimentally accessible \cite{Gu2017}. Likewise, the required phase control $\theta=\pi$ can be readily implemented using existing tunable microwave phase shifters \cite{Pu2010}. These considerations confirm that the optimal parameter regime identified in our study corresponds to experimentally realistic conditions, close to the upper limits of achievable performance, while remaining consistent with stability constraints.

Finally, in Fig.\ref{Hetero}, we present a schematic illustration of the heterodyne detection protocol used to measure the first-order moments and the covariance matrix of a Gaussian state, thereby enabling a direct experimental implementation of the proposed metrological scheme.

\section{CONCLUSIONS}\label{CONCLUSIONS}

In this work, we investigated the quantum multiparameter estimation of the cavity–magnon and magnon–mechanical coupling strengths, $g_{ma}$ and $g_{md}$, in a hybrid microwave cavity–magnon–mechanical system incorporating coherent feedback. Using the QCRB as the ultimate benchmark, we computed the QFIM within the multimode Gaussian formalism and analyzed the corresponding saturation conditions. We further compared the QCRB with the classical Cramér–Rao bound obtained from heterodyne detection, demonstrating that this measurement strategy provides an efficient and experimentally realistic approach to joint parameter estimation.

Our analysis reveals that precise tuning of the cavity and magnon detunings, together with operation at low temperatures, is essential for minimizing estimation errors. The application of a strong microwave drive further enhances the estimation precision—within stability constraints—by increasing the intracavity populations of photons, magnons, and phonons. Notably, dissipative processes can play a constructive role: moderate cavity and magnon dissipation facilitate information extraction through the output field, leading to a reduction of the QCRB, while a reduced mechanical damping rate suppresses thermal noise and enhances coherent excitation exchange among the subsystems.

We also show that optimal estimation performance is achieved through coherent feedback under suitably chosen phase and reflectivity conditions, which effectively suppress cavity noise and improve robustness against thermal fluctuations. From a practical standpoint, heterodyne detection is shown to closely approach the quantum limit over a wide range of parameters, indicating the feasibility of high-precision quantum metrology even under experimentally accessible conditions.

Overall, our results provide clear guidelines for optimizing quantum parameter estimation in cavity–magnon–mechanical platforms and highlight the potential of coherent feedback as a powerful resource for high-precision quantum sensing.

\section*{Declarations} 

{\bf Data Availability Statement:} No data were used for the research described in the article.\\

{\bf Conflict of interest:} The authors declare that they have no known competing financial interests or personal relationships that could have appeared to influence the work reported in this paper.

\end{document}